\documentclass[prd,aps,showpacs,12pt,amssymb,preprintnumbers,amsmath,amsfonts,nofootinbib]{revtex4-1}

\usepackage{latexsym}
\usepackage[latin1]{inputenc}
\usepackage{amsmath}
\usepackage{cancel}
\usepackage{mathtools}

\newcommand{\ba}{\begin{eqnarray}}
\newcommand{\ea}{\end{eqnarray}}
\newcommand{\be}{\begin{equation}}
\newcommand{\ee}{\end{equation}}

\newcommand{\q}{\boldsymbol{q}}
\newcommand{\p}{\boldsymbol{p}}
\newcommand{\bk}{\boldsymbol{k}}
\newcommand{\bv}{\boldsymbol{v}}

\newcommand{\eps}{{\epsilon}}

\usepackage{hyperref}  
\usepackage{dcolumn}
\usepackage[dvips]{graphicx}
\usepackage{color}
\usepackage{slashed}
\usepackage{ulem}
\definecolor{stcol}{rgb}{1,0,1}

\begin{document}
\date{\today}
\title{Accounting for plasma constituent mass effects in heavy fermion energy 
	loss calculations in hot QED and QCD} 

\author{Marc Comadran${}^{1,2,3}$}
\email{mcomadca60@alumnes.ub.edu}

\author{Cristina Manuel${}^{1,2}$}
\email{cmanuel@ice.csic.es}

\author{Stefano Carignano${}^{4}$}
\email{stefano.carignano@bsc.es}

\affiliation{${}^1$Instituto de Ciencias del Espacio (ICE, CSIC) \\
C. Can Magrans s.n., 08193 Cerdanyola del Vall\`es,  Spain 
and 
\\
${}^2$Institut d'Estudis Espacials de Catalunya (IEEC) \\
C. Gran Capit\`a 2-4, Ed. Nexus, 08034 Barcelona, Spain and
}
\affiliation{${}^3$Departament de F\'isica Qu\`antica i Astrof\'isica 
and Institut de Ci\`encies del Cosmos, Universitat de Barcelona (IEEC-UB), Mart\'i i Franqu\'es 1, 08028 Barcelona,  Spain}
\affiliation{${}^4$ Barcelona Supercomputing Center (BSC) \\ 08034 Barcelona, Spain.
}

\begin{abstract}
We evaluate the collisional energy loss of a energetic  fermion with mass $M$ propagating through a hot QED plasma with temperature $T$,  including  mass corrections, that is, 
keeping the mass $m$ of the fermion constituents of the plasma, assuming   $m \ll T \ll M$. We use the bare theory to compute the contribution of hard momentum transfer collisions, and the Braaten-Pisarski resummed theory, amended with small mass corrections, for the contribution of low momentum transfer collisions, and compute the mass corrections at leading logarithmic accuracy in the regime
where the energy of the heavy fermion obeys $E \ll M^2/T$.
We use dimensional regularization to regulate all possible divergences in the computation. If the fermion mass is of order of the soft scale $eT$, where $e$ is the gauge coupling constant, the mass corrections are of the same order as pure perturbative corrections, while they can be substantial for larger values of $m$. We also evaluate the impact of this correction for a QCD plasma.
\end{abstract}
		
\maketitle
		
\section{Introduction}
		
Jet energy loss is a prominent probe for characterizing the properties of matter in heavy-ion collision experiments. Theoretical predictions for the energy loss of an energetic parton produced in the early stages of the collision, which subsequently interacts with the constituents of the produced quark-gluon plasma, can provide invaluable guidance for extracting  properties of the system from experimental data (for recent reviews see eg. \cite{dEnterria:2009xfs,Casalderrey-Solana:2007knd,Majumder:2010qh,Qin:2015srf}).

The first estimate of the collisional energy loss of a heavy fermion in a quark-gluon plasma was made by Bjorken more than 40 years ago
\cite{Bjorken:1982tu}. A naive computation of this quantity is affected by logarithmic infrared (IR) divergences associated to collisions with low momentum transfer.  A good estimate could be done by choosing physical reasonable cutoffs for the momentum transfer. A much more detailed full computation was then carried out by
Braaten and Thoma (BT), first for QED \cite{Braaten:1991jj} and then also for QCD \cite{Braaten:1991we}.  These authors showed how to deal with these divergences consistently, by separating the energy loss computation into two parts, one that would consider high values (or hard) of momentum transfer, or the order of the temperature $T$, and the other with low values (or soft)
of momentum transfer, of order $e T$, where $e$ is the gauge coupling constant. The last should be treated using the hard thermal loop resummed photon propagators, according to the Braaten-Pisarski resummation program \cite{Pisarski:1988vd,Braaten:1989mz}. In the original BT treatment, these two contributions were computed by including an artificial cutoff in momentum transfer to split the two momentum transfer regions, such that when the two contributions are added, the cutoff dependence disappears. The computation by Braaten and Thoma was later on reviewed by Peigne and Peshier \cite{Peigne:2007sd}, correcting the computation for ultrarelativistic
fermions, and further discussing the QCD energy loss in 
\cite{Peigne:2008nd}. These authors  stressed  that the cutoff separating the two different regions should be implemented differently than in Ref.~\cite{Braaten:1991jj}. 

Even if the relevant computations of collisional energy loss are more than 30 years old, there have not been attempts so far to evaluate how they are  perturbatively corrected, see \cite{Ghiglieri:2020dpq}
for the status of perturbation theory for QCD plasmas, for example.
Only recently for QED the perturbative correction to the hard thermal loops (HTL) associated to the photon degrees of freedom has been computed. The perturbative correction arises from both the so called power corrections of the HTL \cite{Manuel:2016wqs,Carignano:2021zhu,  Carignano:2017ovz}, and also from two-loop diagrams \cite{Carignano:2019ofj,Gorda:2022fci,Ekstedt:2023anj}. It has also been noted that the HTLs might be also corrected by effects associated to the (small) mass of the fermions in the plasma \cite{Comadran:2021pkv}. If the mass is of order $\sim e T$, then these mass corrections should be equally important as the genuine perturbative corrections \cite{Comadran:2021pkv}. 

In this article we  evaluate how the collisional energy loss 
 is modified by a small fermion mass of the QED plasma constituents at leading logarithmic accuracy.
We will thus concentrate only in the $t$-channel photon exchange diagram, as the  $s$ and $u$-channel diagrams (Compton scattering) are subleading in the regime we are considering.
 We will assume that there is a good separation of scales, such that $ e T \ll m \ll T$, as if  $m \sim eT$ our results should be corrected by including genuine  perturbative corrections. We will also comment on how the computation should be carried if 
the  fermion mass gets close to $T$. While we provide all the ingredients for such a computation, we will not address such a case here, as it requires a more detailed analysis.

We will use dimensional regularization (DR) to separate the high (or hard) and low (or  soft) momentum transfers parts of the computation. DR is a regularization respectful with gauge invariance. With its use we avoid the ambiguities of how a cutoff is implemented. DR was used in Ref.\cite{Carignano:2021mrn}  to recover the leading logarithmic behavior of Ref.\cite{Braaten:1991jj}. In this paper we also show that the finite term can be recovered using DR, reproducing the same result than the original one in  Ref.\cite{Braaten:1991jj}.

As we will see, the inclusion of a small fermion mass  is non trivial, as new infrared divergences arise in intermediate steps in the computation. 
Those divergences can only be treated consistently with the use of dimensional regularization. This was already apparent in the computation of the small mass corrections to the HTL, see Ref.~\cite{Comadran:2021pkv}, as if the IR is regulated with a cutoff one generates a correction to the photon polarization tensor  that violates gauge invariance. Even if IR divergences arise in the intermediate steps of the mass corrections
to the HTL, the final result is finite, as there is a subtle cancellation of divergences \cite{Comadran:2021pkv}. We witness a similar cancellation of 
the IR divergences associated to the small mass expansion in the hard part of collisional energy loss. After this regularization, only the IR divergence associated to the low momentum transfer remains, which is cancelled with the ultraviolet (UV) divergence to the soft part, yielding a finite result.

We have organized this paper in a way very similar to Ref.~\cite{Braaten:1991jj}, where the computation was carried out with the use a cutoff, emphasizing the points where our treatment is different, and including the corresponding mass corrections. The hard contribution to the collisional energy loss is given in section \ref{Hard}. We provide two different computations of the soft contribution, the first in section \ref{Soft}, which is based on writing the collisional energy loss  as a function of the heavy fermion damping rate, the second one in section \ref{alternative_soft}, based on writing the scattering rate of the collision, using resummed propagators. We present  our final results in section \ref{Final}, where we evaluate the relevance of the mass corrections in the QED collisional energy loss, and comment on the same effect for the QCD plasma. We show in  Appendix \ref{A} some details of the computation of the hard sector, and in Appendix \ref{B}, the integral needed for the computation of soft momentum transfer. We provide in Appendix \ref{C} the form of the polarization tensor that would be needed  for the computation for a generic fermion mass $m$.

We denote four momenta with capital letters, $K^\mu =(k, {\bk})$, and
denote with boldface letters 3 dimensional vectors.
Natural units $\hbar =k_B=c=1$  and metric conventions $g^{\mu \nu} = {\rm diag }(+1,-1,-1,-1)$ are used throughout this manuscript. 

\begin{figure}
	\centering
	\DeclareGraphicsExtensions{.pdf}
	\includegraphics[scale = 1 ]{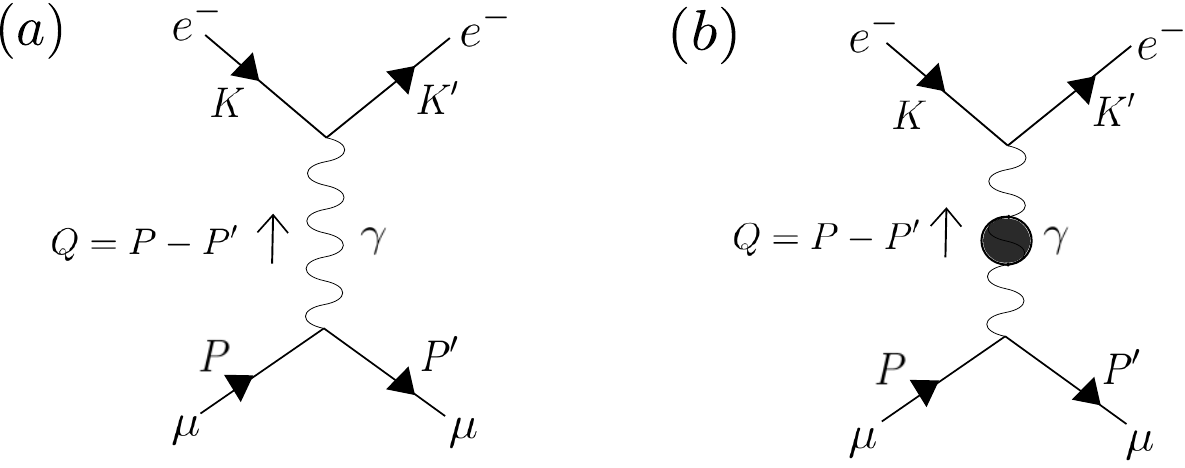}
	\caption{A highly energetic and massive fermion ($\mu$) scatters with the electrons/positrons ($e^-/e^+$) of the medium by the exchange of a virtual photon ($\gamma$). The diagrams with positrons $e^+$ should also be considered (reversing the electron line). $(a)$ Hard contribution. $(b)$ Soft contribution. With the blob representing the resumed propagator.}
	\label{diagram}
\end{figure}

\section{Hard contribution to the collisional energy loss} 
\label{Hard}
The hard contribution to the energy loss from the process $e^{-}\mu \rightarrow e^- \mu$ is depicted diagrammatically in Fig.\hyperref[diagram]{1.(a)}. The diagram with positrons gives exactly the same result, which accounts for a global factor of 2.
We use the following 
notation for the kinematic variables.
The four momentum of the heavy fermion of mass $M$ is
$P^\mu =(E,\p)$, and $K^\mu=(E_k,\bk)$ is the four momentum of the plasma constituents, with mass $m$. We denote with $Q=P-P'=(\omega,\q)=(E-E',\p-\p')$ the momentum of the virtual photon, where we use primed variables for the outgoing particles in the collision. \\
Our starting expression for the computation of the hard contribution to the energy loss in $d$ spatial dimensions is
\be
\begin{gathered}\label{Energy_loss}
	-\dfrac{dE}{dx}\bigg\vert = \dfrac{1}{E} \int \dfrac{d^dp'}{(2\pi)^d 2 E'} \int \dfrac{d^dk}{(2\pi)^d 2 E_k}n_F(E_k) \int \dfrac{d^dk'}{(2\pi)^d 2 E_{k}'}\left[ 1-n_F(E_{k}')\right] \\
	\times (2\pi)^{d+1}\delta^{d+1}(P+K-P'-K')\dfrac{E-E'}{v} \dfrac{1}{2}\sum_{\text{spins}} \vert \mathcal{M}\vert^2 \ ,
\end{gathered}
\ee
where $\mathcal{M}$ is the matrix element associated to the tree-level Feynman diagram. The matrix element is averaged over the spin of the incoming heavy fermion $(\mu)$, and the sum runs over the spins of the involved particles in the process. In Feynman gauge, squaring the amplitude and performing the sum over spins gives
\be
\dfrac{1}{2}\sum_{\text{spins}}\vert \mathcal{M}\vert^2=\dfrac{16e^4\nu^{6-2d}}{t^2}E^2\bigg\lbrace 2(E_k-\bv \cdot \bk)(E_k'-\bv\cdot \bk') +\dfrac{d-1}{2}\dfrac{t^2}{4E^2}+\dfrac{m^2+M^2}{2E^2}t
\bigg\rbrace \ , 
\ee
where $\bv=\p/E$ is the velocity of the heavy fermion and $t=Q^2$.
Here $\nu$ is the renormalization scale that naturally enters in the computation, as 
in changing the spatial dimensions from 3 to $d$ the gauge coupling constant is modified from $e^2$ to $e^2 \nu^{3-d}$. 
Let us stress that no approximation has been used in order to derive the last expression, and we can recover the result for the amplitude squared given in Ref.\cite{Peigne:2007sd} taking the ultra relativistic limit for the electrons/positrons of the plasma. 
Using the delta function in Eq.\eqref{Energy_loss} we can integrate over $\p'$, which yields 
\be \label{Hard_contribution_1}
\begin{gathered}
	-\dfrac{dE}{dx}\bigg\vert^{\text{hard}}= \dfrac{8\pi e^4\nu^{6-2d}}{v} \int \dfrac{d^dk}{(2\pi)^d}n_F(E_k) \int \dfrac{d^dk'}{(2\pi)^d }\left[ 1-n_F(E_{k}')\right]\dfrac{1}{2E'}\delta(\omega+E_k-E_k') \\
	\times  \dfrac{E}{E_k E_k'}\dfrac{\omega}{t^2}\bigg\lbrace 2(E_k-\bv\cdot \bk)(E_k'-\bv\cdot \bk') +\dfrac{d-1}{2}\dfrac{t^2}{4E^2}+\dfrac{m^2+M^2}{2E^2}t\bigg\rbrace \ .
\end{gathered}
\ee
It is convenient to rewrite the remaining delta function of energy conservation in terms of the energy and momentum of the virtual photon
\be \label{delta_function_change_1}
\dfrac{1}{2E'} \delta(\omega+E_k-E_k')=\dfrac{1}{2E}\delta\left( \omega-{\bv}\cdot{\q}-t/(2E)\right) \ .
\ee
As discussed in Refs.\cite{Braaten:1989mz,Peigne:2007sd}, in the Pauli-blocking factor $1-n_F(E_{k}')$ of Eq.\eqref{Hard_contribution_1}, $n_F(E_{k}')$ can be dropped if the energy of the plasma constituents is assumed to be of the order of the temperature i.e $E_k'\sim T$. Indeed, in this regime $t/(2E)\sim T^2/E$ is suppressed inside the delta function of Eq.\eqref{delta_function_change_1} and then the corresponding term in the integrand is odd under the interchange of $\bk$ and $\bk'$, while the measure is even, hence it integrates to zero. Introducing a mass $m$ for the constituents of the plasma does not change any of these assumptions, as long as we do not allow the mass to be higher than the temperature. \\
Introducing the following identity
\be \label{delta_completeness}
1=\int d^dq \delta^d(\q+\bk-\bk')\int d\omega \delta (\omega+E_k-E_{k}') \ ,
\ee
in Eq.\eqref{Hard_contribution_1} we can perform the integration over $\bk'$ using the delta function. The remaining delta function of energy conservation in Eq.\eqref{delta_completeness} can also be written in terms of $\omega$ and $q$ using the relation
\be \label{delta_function_change_2}
\dfrac{1}{2E_{k+q}} \delta (\omega+E_k-E_{k+q})=\delta(t+2\omega E_k-2\bk\cdot\q) \ .
\ee
Applying all these changes we can cast the hard contribution to the energy loss in terms of the energy and momentum transfer variables
\be\label{Hard_contribution_transfer_variables}
\begin{gathered}
	-\dfrac{dE}{dx}\bigg\vert^{\text{hard}}= \dfrac{8\pi e^4\nu^{6-2d}}{v} \int \dfrac{d^dk}{(2\pi)^d}\dfrac{n_F(E_k)}{E_k } \int \dfrac{d^dq}{(2\pi)^d }\int d\omega \delta(t+2\omega E_k-2\bk\cdot\q)\delta\left( \omega-{\bv}\cdot{\q}-t/(2E)\right) \\
	\times \dfrac{\omega}{t^2} \bigg\lbrace 2(E_k-\bv\cdot \bk)(E_{k+q}-\bv\cdot \bk-\omega)+(E_k-\bv\cdot \bk)\dfrac{t}{E}+\dfrac{d-1}{2}\dfrac{t^2}{4E^2}+\dfrac{m^2+M^2}{2E^2}t\bigg\rbrace \ .
\end{gathered}
\ee
If the plasma is macroscopically isotropic, the energy loss should not depend on the particular direction of the velocity of the heavy fermion. Thus, it is common to average \cite{Braaten:1991jj} the last expression over the directions of $\bv$. The required formulas in $d$ spatial dimensions are 
\begin{subequations}
\begin{gather}\label{average_driections_v_a}
	\dfrac{1}{S_d}\int d\Omega_d \delta(\tilde{\omega}-\bv \cdot \q)=\dfrac{1}{L_d}\left( 1-\dfrac{\tilde{\omega}^2}{v^2q^2}\right)^{(d-3)/2}\dfrac{1}{vq}\Theta(v^2q^2-\tilde{\omega}^2) 
	\ , 
    \\\label{average_driections_v_b}\dfrac{1}{S_d}\int d\Omega_d v^i\delta(\tilde{\omega}-\bv \cdot \q)=\dfrac{1}{L_d}\left( 1-\dfrac{\tilde{\omega}^2}{v^2q^2}\right)^{(d-3)/2}\dfrac{1}{vq}\Theta(v^2q^2-\tilde{\omega}^2)\dfrac{\tilde{\omega}}{q}\hat{q}^i
	\ ,
    \\\tag*{}
    \dfrac{1}{S_d}\int d\Omega_d v^iv^j\delta(\tilde{\omega}-\bv \cdot \q)=
    \\\label{average_driections_v_c}
    \dfrac{1}{L_d}\left( 1-\dfrac{\tilde{\omega}^2}{v^2q^2}\right)^{(d-3)/2}\dfrac{1}{vq}\Theta(v^2q^2-\tilde{\omega}^2)\left( \dfrac{v^2q^2-\tilde{\omega}^2}{(d-1)q^2}\delta^{ij}+\dfrac{d\tilde{\omega}^2-v^2q^2}{(d-1)q^2}\hat{q}^i\hat{q}^j\right) \ ,
\end{gather}
\end{subequations}
where $\Theta(x)$ is the Heaviside step function, we define $\tilde{\omega}=\omega-t/(2E)$ and use the notation $S_d= \int d\Omega_d= {2\pi^{d/2}}/{\Gamma(d/2)}$ for the surface of a $d$-sphere and  $L_d=\Gamma(\frac{d-1}{2})\sqrt{\pi}/\Gamma(\frac{d}{2})$ . \\
After the average over the directions of $\bv$, we can further simplify Eq.\eqref{Hard_contribution_transfer_variables} introducing the explicit form of the measures in $d$ dimensions and using the remaining delta function to perform the integration over the angle $\theta_{\bk,\q}\equiv \theta$. The result may be written as
\be \label{angular_integration}
\begin{gathered}
	\int_{-1}^1 d(\cos\theta)(\sin\theta)^{d-3} \delta(t+2\omega E_k-2kq \cos\theta)=
	\\
	= \dfrac{1}{2kq} \Theta\left( \sqrt{(q-k)^2+m^2}\leq \vert\omega+E_k \vert\leq \sqrt{(q+k)^2+m^2}\right)   \left(   1-\left[ \dfrac{t}{2kq}+\dfrac{\omega E_k}{kq}\right]   ^2 \right)  ^{(d-3)/2} \ .
\end{gathered}
\ee
Performing these changes, the hard contribution to the energy loss reads
\be\label{Hard_contribution_in_d}
\begin{gathered}
	-\dfrac{dE}{dx}\bigg\vert^{\text{hard}}= \dfrac{2 e^4\nu^{6-2d}}{v^2} C_d \int_0^\infty dkk^{d-2}\dfrac{n_F(E_k)}{E_k} \int_0^\infty dqq^{d-5}
	\\
	\times\int d\omega \omega  \Theta\left( \sqrt{(q-k)^2+m^2}\leq \vert\omega+E_k \vert\leq \sqrt{(q+k)^2+m^2}\right) \Theta(v^2q^2-\tilde{\omega}^2)  
	\\
	\times  \dfrac{q^2}{t^2} 
	\bigg\lbrace2E_k^2-2\dfrac{E_k\tilde{\omega}}{q^2}(t+2\omega E_k)+ \dfrac{2}{d-1}\left[ \dfrac{v^2q^2-\tilde{\omega}^2}{q^2}k^2+\dfrac{d\tilde{\omega}^2-v^2q^2}{4q^4}(t+2\omega E_k)^2\right] 
	\\
	+\left( E_k-\tilde{\omega}\dfrac{t+2\omega E_k}{2q^2}\right) \dfrac{t}{E}+\dfrac{d-1}{2}\dfrac{t^2}{4E^2}+\dfrac{m^2+M^2}{2E^2}t\bigg\rbrace \times \mathcal{K}_d(\omega,q,k) \ .
\end{gathered}
\ee
Here we collected the numerical factors and functions arising from the $d$ dimensional integration measures in          %
\be \label{constant}
C_d =\frac{2^{3-d}}{\Gamma(\frac{d-1}{2})^2(2\pi)^d} \ ,
\ee
and
\be \label{function}
\mathcal{K}_d(\omega,q,k)=\left( 1-\dfrac{\tilde{\omega}^2}{v^2q^2}\right) ^{\frac{(d-3)}{2}} \times
\left(    1-\left[  \dfrac{t}{2kq}+\dfrac{\omega E_k}{kq}\right]   ^2 \right)  ^{\frac{(d-3)}{2}} \ ,
\ee
respectively. The theta functions in Eq.\eqref{Hard_contribution_in_d} affect the integration boundaries of the energy and momentum transfer integrals. Let us discuss how introducing a mass $m$ for the plasma constituents modifies the boundaries given in Ref.\cite{Peigne:2007sd}. From the theta function $\Theta(v^2q^2-\tilde{\omega}^2)$ we get the boundaries for the energy transfer $\omega_{\pm}(q)= E-\sqrt{E^2+q^2\mp 2Evq}$, then
taking into account $-q\leq \omega_-(q)\leq \omega_+(q)\leq q$ it can be shown that
\be \label{theta_constraints}
\begin{gathered}
	\Theta\left( \sqrt{(q-k)^2+m^2}\leq \vert\omega+E_k \vert\leq \sqrt{(q+k)^2+m^2}\right) \Theta(\omega_-\leq \omega \leq \omega_+)=
	\\
	=
	\Theta(0\leq q\leq q_{\text{in}})\Theta(\omega_-\leq \omega \leq \omega_+)
	+\Theta(q_{\text{in}}\leq q\leq q_{\text{max}})\Theta(\omega_{\text{min}} \leq \omega \leq \omega_+) \ .
\end{gathered}
\ee
Being $\omega_{\text{min}}(q)=-E_k+\sqrt{(q-k)^2+m^2}$. In addition, the limits of integration for the momentum transfer $q_{\text{in}}$ and $q_{\text{max}}$ are obtained solving the equations
\be
\begin{gathered}
	\vert\omega_-(q_{\text{in}}) +E_k\vert = \sqrt{(q_{\text{in}}-k)^2+m^2} \ 
	\quad \text{and} \quad
	\vert\omega_+(q_{\text{max}}) +E_k\vert = \sqrt{(q_{\text{max}}-k)^2+m^2} \ ,
\end{gathered}
\ee
which gives
\be \label{qin_qmax}
\begin{gathered}
	q_{\text{in}}=\dfrac{2E^2(k- vE_k )+2EE_k(k-vE_k)}{E^2(1-v^2)+2E(E_k-vk)+m^2} \ 
	\quad \text{and} \quad
	q_{\text{max}}=\dfrac{2E^2(k+v E_k )-2EE_k(k-vE_k)}{E^2(1-v^2)+2E(E_k+vk)+m^2} \ ,
\end{gathered}
\ee
respectively. Thus, according to Eq.\eqref{theta_constraints}, the integration over energy and momentum transfer is separated into two regions
\be
\begin{gathered}\label{boundaries_hard}
	\int_0^\infty dq \int_{\omega_-(q)}^{\omega_+(q)}d\omega \Theta\left( \sqrt{(q-k)^2+m^2}\leq \vert\omega+E_k \vert\leq \sqrt{(q+k)^2+m^2}\right) \longrightarrow
	\\
	\longrightarrow \int_0^{q_{\text{in}}}dq\int_{\omega_-(q)}^{\omega_+(q)}d\omega + \int^{q_{max}}_{q_{\text{in}}}dq\int_{\omega_{\text{min}}(q)}^{\omega_+(q)}d\omega \ . 
\end{gathered}
\ee
Now we re-express $\tilde{\omega}=\omega-t/(2E)$ in Eq.\eqref{Hard_contribution_in_d} and organize all terms in inverse powers of $E$. Then the hard contribution to the energy loss may be written as
\be
\label{Hard_contribution_powers_1/E}
\begin{gathered}
	-\dfrac{dE}{dx}\bigg\vert^{\text{hard}}= \dfrac{2 e^4\nu^{6-2d}}{v^2} C_d \int_0^\infty dkk^{d-2}\dfrac{n_F(E_k)}{E_k}
	\left\lbrace \int_0^{q_ {\text{in}}}dq q^{d-5}\int_{\omega_-(q)}^{\omega_+(q)}d\omega \omega + \int^{q_{max}}_{q_ {\text{in}}}dqq^{d-5} \int_{\omega_{\text{min}}(q)}^{\omega_+(q)}d\omega \omega \right\rbrace 
	\\
	\times 
	\bigg\lbrace  \dfrac{3\omega^2-v^2q^2}{4q^2}
	+\dfrac{3E_k(E_k+\omega)}{q^2}
	+\left[ \dfrac{m^2q^2}{t^2}
	+\dfrac{E_k(E_k+\omega)}{t}+\dfrac{q^2}{2t}\right] (1-v^2)
	+\dfrac{m^2}{t}
	\\
	-\dfrac{\omega\left[ 12E_k(E_k+\omega)+3\omega^2-q^2\right] }{4q^2E}-\dfrac{\omega m^2}{tE}  
	+\dfrac{4E_k(E_k+\omega)(3\omega^2-q^2)+3(\omega^4+q^4)-2\omega^2q^2}{16q^2E^2}+\dfrac{m^2}{4}\dfrac{\omega^2+q^2}{tE^2}
	\\
	-\dfrac{d-3}{2}\bigg[ \dfrac{1}{4}+\dfrac{m^2q^2}{t^2}+\dfrac{E_k(E_k+\omega)}{t}\bigg]\left( \dfrac{\omega^2-v^2q^2}{q^2}+\dfrac{t^2}{4q^2E^2}-\dfrac{\omega t}{q^2E}\right)   +\dfrac{d-3}{2}\dfrac{q^2}{4E^2}
	\bigg\rbrace
	\times \mathcal{K}_d(\omega,q,k)\ .
\end{gathered}
\ee
We expanded the terms inside the curly brackets of Eq.\eqref{Hard_contribution_in_d} for $d\rightarrow 3$, keeping only pieces proportional to $d-3$, as they are needed for the computation of the finite pieces of the energy loss.
Let us recall which assumptions are necessary to extract the leading order pieces to the energy loss, which are of order $\sim e^4T^2$, for a detailed discussion see Ref.\cite{Peigne:2007sd}.
In thermal equilibrium, the energy of most plasma constituents is of the order of temperature i.e $E_k\sim T$. In addition, we assume that the energy of the incoming heavy fermion is much larger than the temperature. This gives the hierarchy $E_k \sim T \ll E$. Furthermore, the integration boundaries for the momentum transfer can be simplified in the limit $E\ll M^2/T$ as
\be
\begin{gathered} \label{boundaries_E << M^2/T}
	q_{\text{in}}\approx 2\dfrac{k-vE_k }{1-v^2} ,
	\qquad \text{and} \qquad
	q_{\text{max}}\approx 2\dfrac{k+vE_k }{1-v^2}\ .
\end{gathered} 
\ee
Consequently, the momentum transfer is much smaller than the energy of the heavy fermion $q\sim E^2 /(M^2/T) \ll E$, so we can assume that it is of the order of the temperature $q\sim T \ll E$. In addition, in the region $0 \leq q\leq q_\text{in}$ the integration boundaries for the energy transfer can be approximated as
$\omega_{\pm}(q)\approx \pm vq $, so we conclude that the energy transfer is also of the order of the temperature  $\omega\sim T $. 
Now it can be easily seen that the terms which are not suppressed by powers of $E$ in Eq.\eqref{Hard_contribution_powers_1/E} are of order $e^4T^2$ while all other terms are of order $e^4T^3/E$ and  $e^4T^3/E^2$ so they can be ignored in the regime $T\ll E$. Taking this into account and moving to the dimensionless variable $x=\omega/q$, we reach to
\be\label{Hard_contribution_in_d_leading_order}
\begin{gathered}
	-\dfrac{dE}{dx}\bigg\vert^{\text{hard}}= \dfrac{2 e^4\nu^{6-2d}}{v^2} C_d \int_0^\infty dkk^{d-2}\dfrac{n_F(E_k)}{E_k} \left\lbrace \int_0^{2\frac{k-vE_k }{1-v^2}}dq q^{d-5}\int_{-v}^{v}dx x + \int^{q_{max}}_{2\frac{k-vE_k }{1-v^2}}dqq^{d-5} \int_{x_{\text{min}}(q)}^{x_+(q)}dx x \right\rbrace 
	\\
	\times\bigg\lbrace  
	(E_k^2+qxE_k )\left( 3+\dfrac{1-v^2}{x^2-1}\right) +m^2\dfrac{x^2-v^2}{(x^2-1)^2}   +q^2\left( \dfrac{3x^2-v^2}{4}+\dfrac{1-v^2}{2(x^2-1)}\right)
	\\
	-\dfrac{d-3}{2}\bigg[(E_k^2+qxE_k)\dfrac{ x^2-v^2}{x^2-1}+ m^2\dfrac{ x^2-v^2}{(x^2-1)^2}+\dfrac{q^2}{4}( x^2-v^2)\bigg]
	\bigg\rbrace \times \mathcal{K}_d(x,q,k) \ .
\end{gathered}
\ee
The momentum transfer integrals in the integration region $0\leq q \leq q_{\text{in}}$ contain an IR divergence, while in the region $q_{\text{in}}\leq q \leq q_{\text{max}}$ the momentum transfer integrals are finite.
We focus now on the dominant IR divergent region, and use DR to regularize possible divergences. 
The terms $\sim q^{d-4}$ in the integrand of  Eq.\eqref{Hard_contribution_in_d_leading_order} yield  the relevant IR divergent integral in momentum transfer, which we evaluate (see Appendix \ref{A}) 
\be \label{q_integral_hard_IR}
\begin{gathered}
	\nu^{3-d}\int_0^{2\frac{k-vE_k }{1-v^2}}dq q^{d-4} \left( 1-\left[ \dfrac{q(x^2-1)}{2k}+\dfrac{x E_k}{k}\right]^2\right)^{\frac{(d-3)}{2}}  
	= 
	\\
	=
	\dfrac{1}{d-3}\left( 2\dfrac{k-vE_k}{(1-v^2)\nu}\right)^{d-3}\left( 1-\dfrac{x^2E_k^2}{k^2}\right)^{\frac{(d-3)}{2}} +  \mathcal{O}\left(d-3\right) \ .
\end{gathered}  
\ee
The momentum transfer integral for the terms $\sim q^{d-5}$ in Eq.\eqref{Hard_contribution_in_d_leading_order} is free of divergences, but is necessary to reproduce the finite pieces of Ref.\cite{Braaten:1991jj} as well as the mass corrections to the pole computed in this manuscript. The required momentum transfer integral is
\be \label{q_integral_hard_finite}
\begin{gathered}
	\nu^{3-d}\int_0^{2\frac{k-vE_k }{1-v^2}}dq q^{d-5} \left( 1-\left[ \dfrac{q(x^2-1)}{2k}+\dfrac{x E_k}{k}\right]^2\right)^{\frac{(d-3)}{2}}  
	= 
	\\
	=
	\left( 1-\dfrac{x^2E_k^2}{k^2}\right)^{\frac{(d-3)}{2}}\left( 2\dfrac{k-vE_k}{(1-v^2)\nu} \right) ^{d-3} \left\lbrace \dfrac{1}{d-4} \left( 2\dfrac{k-vE_k}{1-v^2}\right)   +  x \dfrac{ E_k}{2k^2}\dfrac{1-x^2}{1-x^2E_k^2/k^2} +\mathcal{O}\left( d-3 \right) \right\rbrace \ .
\end{gathered}  
\ee
Then, the first term inside the curly brackets above does not contribute to the pole, as it would vanish due to antisymmetry in $x$, while the second term does not and thus must be taken into account\footnote{Notice that as $d\rightarrow 3$ the measure of \eqref{Hard_contribution_in_d_leading_order} is symmetric in $x$ in the region $0<q<q_{\text{in}}$.}.
The remaining terms, those $\sim q^{d-3}$ of Eq.\eqref{Hard_contribution_in_d_leading_order} are not necessary for the evaluation of the hard contribution in the region $0<q<q_{\text{in}}$. We give more details on this last statement and the computation of the momentum transfer integrals in Appendix \ref{A}. Using Eq.\eqref{q_integral_hard_finite} and Eq.\eqref{q_integral_hard_IR} we may write the hard contribution to the energy loss as
\be \label{Hard_contribution_after_q_integral}
\begin{gathered}
	-\dfrac{dE}{dx}\bigg\vert^{\text{hard}}\approx \dfrac{2 e^4 \nu^{3-d}}{v^2} C_{d} \int_0^\infty dkk^{d-2}n_F(E_k)
	\int_{-v}^{v}dx  \left( 1-{x^2}/{v^2}\right) ^{(d-3)/2} \left( 1-{x^2E_k^2}/{k^2}\right)^{(d-3)/2}
	\\
	\times\dfrac{1}{d-3} \left(2\frac{k-vE_k }{(1-v^2)\nu} \right)^{d-3}\bigg\lbrace 
	3x^2+\dfrac{x^2(1-v^2)}{x^2-1}+ \dfrac{d-3}{2}\bigg[ 
	\dfrac{E_k^2}{k^2}\dfrac{1-x^2}{1-x^2E_k^2/k^2}\left( 3x^2+\dfrac{x^2(1-v^2)}{x^2-1}\right) 
	\\
	+\dfrac{m^2}{k^2}\dfrac{1-x^2}{1-x^2E_k^2/k^2}\dfrac{x^2(x^2-v^2)}{(x^2-1)^2}-\dfrac{x^2(x^2-v^2)}{x^2-1} \bigg]+\mathcal{O}\left[(d-3)^2 \right] \bigg\rbrace 
	\ .
\end{gathered}
\ee
Let us recall that we have not yet made any assumption for the mass $m$ of the constituents of the plasma. Hence, the above expression may be valid for arbitrary mass $m$ as long as it does not surpass the plasma temperature. Though, in this work we assume that the mass of the plasma constituents is smaller than the temperature of the thermal bath $m \ll  T $, which allows us to expand Eq.\eqref{Hard_contribution_after_q_integral} for small $m$. Performing such expansion produces new pieces, some of them potentially divergent for $k\rightarrow 0$. This is the reason why we kept explicit $\mathcal{O}(d-3)$ pieces in the integrand of Eq.\eqref{Hard_contribution_after_q_integral}, since those pieces, after expanding for small mass $m$ produce IR divergent terms i.e $\sim 1/(d-3)$, thus giving a contribution to the pole. Note however, that if we computed the $k$ integrals in Eq.\eqref{Hard_contribution_after_q_integral} for a generic mass $m$, all $\mathcal{O}(d-3)$ pieces would yield only finite contributions, because in this scenario the $k$ integrals are free of divergences (the mass $m$ acts as a lower cut-off). 
That generic case could be considered if the Braaten-Pisarski resummation program is generalized for generic values of the fermion mass. We comment on how this should be done in the remaining part of the paper. 
\\ 
Setting $m=0$ in Eq.\eqref{Hard_contribution_after_q_integral} gives the leading order pieces of the hard contribution to the energy loss 
\be\label{Hard_leading_order_m=0}
\begin{gathered}
	-\dfrac{dE}{dx}\bigg\vert^{\text{hard}}_{m=0}\approx \dfrac{2 e^4\nu^{3-d}}{v^2} C_{d} \int_0^\infty dkk^{d-2} n_F(k)
	\int_{-v}^{v}dx  \left( 1-{x^2}/{v^2}\right) ^{(d-3)/2} \left( 1-x^2\right) ^{(d-3)/2} 
	\\
	\times \dfrac{1}{d-3} \left(\frac{2k}{(1+v)\nu}\right)^{d-3}\left\lbrace  3x^2+\dfrac{x^2(1-v^2)}{x^2-1}+(d-3) x^2\right\rbrace \ . 
\end{gathered}
\ee
All remaining integrals are finite and can be computed analytically, they are given in Appendix \ref{A}. Using the expressions derived there we find
\be
\label{hard_final_m=0}
\begin{gathered}
	-\dfrac{dE}{dx}\bigg\vert^{\text{hard}}_{m=0}\approx \dfrac{e^2 m^2_{D,3+2\eps}}{16\pi}\bigg\lbrace \left( \dfrac{1}{\eps}+\ln \dfrac{4T^2}{(1+v)^2\overline{\nu}^2}\right) \left( \dfrac{1}{v}-\dfrac{1-v^2}{2v^2}\ln\dfrac{1+v}{1-v}\right)+\dfrac{2v}{3} +\dfrac{P(v)}{v^2}\bigg\rbrace
	\\
	+\dfrac{e^2 m^2_{D,3+2\eps}}{8\pi} \left( \dfrac{1}{v}-\dfrac{1-v^2}{2v^2}\ln\dfrac{1+v}{1-v}\right) \left\lbrace 1-\gamma_E+\dfrac{\zeta'(2)}{\zeta(2)}+\ln 2\right\rbrace  \ .
\end{gathered}
\ee
Where we defined the Debye mass in $d$ dimensions
 \be\label{Debye_mass_in_d}
\begin{gathered}
	m_{D,d}^2\equiv 16e^2\nu^{3-d}F_d \int_{0}^{\infty} dk\ k^{d-2} n_F(k) \ ,
\end{gathered}
\ee
also $\overline{\nu}^2=4\pi e^{-\gamma_E}\nu^2$,  being $\gamma_E$ the Euler-Mascheroni constant and $\zeta(x)$ is the Riemann zeta function.  Furthermore, we conveniently defined the quantity
\be \label{P(v)}
P(v)=\int_{-v}^{v}dx \,x^2\left( 
1+\dfrac{1}{2}\dfrac{x^2-v^2}{x^2-1}
\right)  \ln\left[ (1-x^2)(1-x^2/v^2)\right] \ .
\ee
Remarkably, regularizing the momentum transfer integrals with DR instead of a cut-off, as was done in references \cite{Braaten:1991jj} and \cite{Peigne:2007sd}, 
produces the same logarithmic dependence in $\overline{\nu}$ that the one found out with a cutoff $q^*$, but it generates extra finite pieces that we have partially absorbed in our definition of $m_{D,3+2\eps}^2$, the function $P(v)$ and the term $2v/3$. However, the very same extra pieces with opposite sign are generated in the soft contribution to the energy loss, and ultimately they cancel when both contributions are added. 
\\
For completeness, we should also include in Eq.\eqref{hard_final_m=0} the finite pieces generated in the momentum transfer region $q_{\text{in}}<q<q_{\text{max}}$. Since the $q$ integrals are finite in that region, there is no need of regularization and we can compute them in $d=3$, reaching the same result of Ref.\cite{Braaten:1991jj}.
\\
We turn now our attention in computing small mass i.e $m \ll T$ corrections to Eq.\eqref{hard_final_m=0}. The scale of the IR divergence in momentum transfer of Eq.\eqref{Hard_contribution_after_q_integral} should be approximated as
\be
\left(2\frac{k-vE_k }{(1-v^2)\nu} \right)^{d-3}\approx \left(\frac{2k}{(1+v)\nu}\right)^{d-3} \ , \ 
\ee
as we concentrate in the IR divergent term in momentum transfer.
 Using the following small $m$ expansions
\begin{subequations}
\begin{gather}
	n_F(E_k)=n_F(k)+\dfrac{m^2}{2k}\dfrac{dn_F}{dk}+\mathcal{O}(m^4) \ ,
	\\
	\dfrac{E_k^2}{k^2}\dfrac{1-x^2}{1-x^2E_k^2/k^2}=1+\dfrac{m^2}{k^2}\dfrac{1}{1-x^2}+\mathcal{O}(m^4) \ ,
	\\
	\left( 1-{x^2E_k^2}/{k^2}\right)^{(d-3)/2}=\left( 1-x^2\right) ^{(d-3)/2}\left( 1-\dfrac{d-3}{2} \dfrac{m^2}{k^2}\dfrac{x^2}{1-x^2}\right) +\mathcal{O}(m^4) \ ,
\end{gather}
\end{subequations}
and keeping only terms up to $\mathcal{O}(m^2)$ in Eq.\eqref{Hard_contribution_after_q_integral} we find
\be \label{Hard_contribution_m^2_after_q_integrals}
\begin{gathered}
	-\dfrac{dE}{dx}\bigg\vert^{\text{hard}}_{m^2}\approx \dfrac{2 e^4m^2\nu^{3-d}}{v^2} C_{d}  \int_0^\infty dkk^{d-3} \int_{-v}^{v}dx  \left( 1-{x^2}/{v^2}\right) ^{(d-3)/2} \left( 1-x^2\right)^{(d-3)/2} 
	\\
	\times 
	\dfrac{1}{d-3} \left(\frac{2k}{(1+v)\nu}\right)^{d-3} 
	\bigg\lbrace 
	\dfrac{1}{2}\dfrac{dn_F}{dk} \left(3x^2+\dfrac{x^2(1-v^2)}{x^2-1} \right) 
	\\
	+\dfrac{d-3}{2} \dfrac{n_F(k)}{k}\left( 3x^2+\dfrac{x^2(1-v^2)}{x^2-1}
	+\dfrac{x^2(x^2-v^2)}{(x^2-1)^2}\right)
	+\mathcal{O}\left[(d-3)^2 \right]  \bigg\rbrace 
	\ .
\end{gathered}
\ee
The momentum integral $\sim n_F(k)/k$ of the above expression contains an extra IR divergence for low $k$, which appears because of the small $m$ expansion. These are the same sort of IR divergences that appear in the computation of the small $m$ corrections to the HTL, which however cancel after performing the angular integrals, see Ref.\cite{Comadran:2021pkv}. A similar situation happens now here. After the complete computation,  only  the IR divergence associated to the low momentum transfer integral survives.
 We discuss this issue in detail in Appendix \ref{A}, where we derive the relevant IR energy integral
\be\label{Integral_k_hard_m^2_II}
\begin{gathered}
	e^4 m^2\nu^{3-d}C_{d}\int_0^{\infty} dk \ k^{d-4} \left(\frac{2k}{(1+v)\nu}\right)^{d-3}n_F(k)=
	\\
	=\dfrac{e^2}{16\pi}\dfrac{e^2m^2}{2\pi^2}\left\lbrace \dfrac{1}{\eps}+\left( -2\gamma_E+2\ln\pi+\ln \dfrac{T^2}{\overline{\nu}^2}+\ln \dfrac{4T^2}{(1+v)^2\overline{\nu}^2}\right)+\mathcal{O}(\eps) \right\rbrace \ .
\end{gathered}
\ee
The logarithm $\ln(T^2/\overline{\nu}^2)$ above is cancelled when the soft contribution is added (see Sec.\hyperref[alternative_soft]{IV}). Also, we can ignore the finite pieces inside the parenthesis of the second line, since we are just interested in extracting the leading logarithmic behaviour. Then, plugging the results Eq.\eqref{Integral_k_hard_m^2_I} and Eq.\eqref{Integral_k_hard_m^2_II} into Eq.\eqref{Hard_contribution_m^2_after_q_integrals}, we note that the pole of the $k$-integral in Eq.\eqref{Integral_k_hard_m^2_II} vanishes. In addition, we see that the remaining necessary integral in energy transfer is
\be\label{Angular_integral_parte_hard_m^2}
\begin{gathered}
	\int_{-v}^{v}dx  \left( 1-{x^2}/{v^2}\right) ^{(d-3)/2} \left( 1-x^2\right)^{(d-3)/2} \dfrac{x^2(x^2-v^2)}{(x^2-1)^2}=
	\\
	=v^2\left\lbrace \left( \dfrac{3}{v}-\dfrac{v^2-3}{2v^2}\ln\dfrac{1+v}{1-v}\right) +\mathcal{O}(\eps)\right\rbrace \ .
\end{gathered}
\ee
Collecting these results we can write down the leading mass correction to the hard contribution
\be\label{Hard_contribution_m^2_result}
\begin{gathered}
	-\dfrac{dE}{dx}\bigg\vert_{m^2}^{\text{hard}}\approx
	\dfrac{e^2}{16\pi}\dfrac{e^2m^2}{2\pi^2}
	\left( \dfrac{1}{\eps}+\ln \dfrac{4T^2}{(1+v)^2\overline{\nu}^2}\right) \left(  \dfrac{3}{v}-\dfrac{v^2-3}{2v^2}\ln\dfrac{1+v}{1-v}\right) 
	\ .
\end{gathered}
\ee
We recall here that in this manuscript we computed leading order mass corrections to the energy loss at logarithmic accuracy, and we have not included the computation of all finite pieces, which would require a more involved analysis. For instance, we would need to consider the explicit $\mathcal{O}\left[(d-3)^2 \right]$ pieces that we have ignored in Eq.\eqref{Hard_contribution_m^2_after_q_integrals} together with the finite pieces generated in the region $q_{\text{in}}\leq q \leq q_{\text{max}}$.
		
\section{Soft contribution to the collisional energy loss} 
\label{Soft}
\begin{figure}
	\centering
	\DeclareGraphicsExtensions{.pdf}
	\includegraphics[scale = 1.5 ]{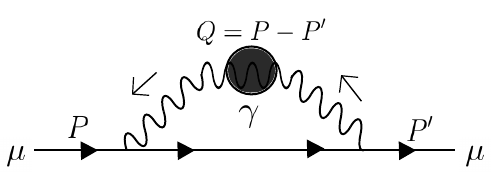}
	\caption{A highly energetic and massive fermion ($\mu$) self-interacts by the exchange of a virtual (resummed) photon ($\gamma$).}
	\label{self_energy}
\end{figure}
In this section we will compute the soft contribution to the energy loss. We will  derive the soft contribution trough the computation of the damping rate of 
the heavy fermion traversing the QED plasma. This requires to compute the imaginary part of the heavy fermion self-energy, and relate the collisional energy-loss to this damping rate, see for example Ref.\cite{Braaten:1991jj}. 
Generalizations in $d$ dimensions of the energy loss formula for the soft contribution  can be found  in Ref.\cite{Carignano:2021mrn}. The final formula appears in terms of the resummed photon propagators.  The mass corrections we are aiming to compute only enter in the
resummed photon propagators. \\
Our starting expression is in Coulomb gauge (see Appendix A of Ref.~\cite{Carignano:2021mrn})
\be \label{Soft_contribution_start}
\begin{gathered}
	-\dfrac{dE}{dx}\bigg\vert^{\text{soft}}=\dfrac{e^2\nu^{3-d}}{v^2}F_d \int_0^{\infty}  dq q^{d-1} \int_{-v}^{v} dx (1-x^2/v^2)^{(d-3)/2} \left[ 1+n_B(qx)\right] 
	\\
	\times (qx)\Big( \rho_{L}^d(qx,q)+(v^2-x^2) \rho_{T}^d(qx,q)\Big) \ .
\end{gathered}
\ee
Here $\rho_{L/T}^d(qx,q)$ stand for the 
longitudinal/transverse HTL photon spectral functions 
in $d$ dimensions, respectively, $F_d=2^{1-d}\pi^{-(d+1)/2}/\Gamma\left( \frac{d-1}{2}\right)$ is the constant that arises from the momentum transfer measure in $d$ dimensions and we used the dimensionless variable $x=q_0/q$. The longitudinal and transverse spectral functions can  be obtained using their relation with the imaginary part of the (resummed) longitudinal and transverse retarded propagators,
i.e $\rho_{L/T}^d(q_0,q)=2\,\text{Im}\, \Delta^d_{L/T}(q_0+i 0^+,q)$ \cite{LeBellac}.
Note that this definition of the spectral function differs by a factor $ (- 2 \pi)$ to the one used in Ref.\cite{Braaten:1991jj}.
In terms of   the longitudinal/transverse retarded polarization tensors
$\Pi_d^{L/R}$ in $d$ dimensions one can write
\begin{subequations} \label{Spectral_functions_self_energies}
\begin{gather}
	\rho^{L}_d(q_0,q)=-\dfrac{2 \,\text{Im}\Pi^{L}_d(q_0,q)}{\left[ q^2-\text{Re}\Pi^{L}_d(q_0,q)\right] ^2+\left[ \text{Im}\Pi^{L}_d(q_0,q)\right]^2  } \ ,
	\\
	\rho^{T}_d(q_0,q)=-\dfrac{2\,\text{Im}\Pi^{T}_d(q_0,q)}{\left[ q_0^2 -q^2-\text{Re}\Pi^{T}_d(q_0,q)\right] ^2+\left[ \text{Im}\Pi^{T}_d(q_0,q)\right]^2 } \ .
\end{gather}
\end{subequations} 
To be able to compute mass corrections to the soft contribution of the energy loss, we will need the mass corrections to the HTL polarization tensor, which were computed in Ref.\cite{Comadran:2021pkv}. In the presence of a small mass  $m \ll T$, the longitudinal and transverse components of the  retarded polarization read 
\begin{subequations}
\begin{gather}
\Pi^{L}_{3+2\eps}(q_0,q)=-m_{D,3+2\eps}^2\left(    1-\dfrac{x}{2}\ln\dfrac{x+1}{x-1} \right)  -\dfrac{m^2e^2}{2\pi^2}\dfrac{1}{x^2-1}+ \mathcal{O}(\eps) \ ,
\\
\Pi^T_{3+2\eps}(q_0,q)=m_{D,3+2\eps}^2\dfrac{x^2}{2}\left( 1-\left(1-\dfrac{1}{x^2} \right)\dfrac{x}{2} \ln \dfrac{x+1}{x-1} \right) -\dfrac{m^2e^2}{2\pi^2} \dfrac{x}{2} \ln \dfrac{x+1}{x-1}+ \mathcal{O}(\eps) \ ,
\end{gather}
\end{subequations}
where $x =(q_0+ i0^+)/q$, and
$m_{D,3+2\eps}^{2}$ is the Debye mass in $d=3+2\eps$ dimensions defined in Eq.\eqref{Debye_mass_in_d}. As discussed in Ref.\cite{Comadran:2021pkv}, when the mass of the fermionic particles obeys $eT \ll m \ll T$ the mass corrections are dominant in comparison 
to the perturbative corrections.
Then, the longitudinal and transverse spectral function including mass corrections read
\be\label{spectral_L}
\rho^{L}_{3+2\eps}(qx,q)=2\pi\dfrac{  (1+\eps) m_{D,3+2\eps}^2 x\, \Theta(1-x^2)(1-x^2)^{\eps}}{2\left(  q^2+m_D^2Q_1(x)-\dfrac{m^2e^2}{2\pi^2}\dfrac{1}{1-x^2}\right)  ^2+\dfrac{\pi^2 x^2}{2}m_D^4 }+\mathcal{O}(\eps) \ ,
\ee
and 
\be\label{spectral_T}
\rho^{T}_{3+2\eps}(qx,q)=\dfrac{2\pi}{1-x^2}\dfrac{ \left(  m_{D,3+2\eps}^2-\dfrac{m^2e^2}{\pi^2}\dfrac{1}{1-x^2}\right) x \, \Theta(1-x^2)(1-x^2)^{\eps}}{\left(   2q^2+m_D^2Q_2(x)-\dfrac{m^2e^2}{2\pi^2}Q_3(x)\right)  ^2+\dfrac{\pi^2 x^2}{4} \left(  m_D^2-\dfrac{m^2e^2}{\pi^2}\dfrac{1}{1-x^2}\right)  ^2}+\mathcal{O}(\eps).
\ee
respectively.  We kept the Debye mass in $d=3+2\eps$ dimensions of the numerators unexpanded for small $\eps$ in the spectral functions for convenience.
 Furthermore, we did not include $\mathcal{O}(\eps)$ pieces arising from the denominator, since those pieces do not generate UV divergences when expanded for small $\eps$, and thus vanish in the limit $\eps\rightarrow 0$. Lastly, we also did not include $\mathcal{O}(\eps)$ pieces coming from the mass corrections to the HTL, although they would be needed in order to determine the mass correction to the soft contribution of the energy loss beyond logarithmic accuracy.
We also introduced, to shorten the notation, the functions
\be\label{Legendre_functions}
Q_1(x)=1-\dfrac{x}{2}\ln \bigg\vert \dfrac{x+1}{x-1}\bigg\vert \ , \quad Q_2(x)=\dfrac{1}{1-x^2}-Q_1(x) \ , \quad Q_3(x)=\dfrac{x}{x^2-1}\ln \bigg\vert \dfrac{x+1}{x-1}\bigg\vert\ .
\ee
Due to the symmetries of the integrand, we can replace $1+n_B(qx)$ in Eq.\eqref{Soft_contribution_start} by its even part, which is just 1/2. The spectral functions should also be expanded for small mass $m$, however, the pieces generated are subleading corrections, so we can ignore them.  Taking these remarks into account, we plug the spectral functions of Eqs.\eqref{spectral_L}-\eqref{spectral_T} in Eq.\eqref{Soft_contribution_start} reaching to
\be \label{Soft_contribution_intermediate_step}
\begin{gathered}
	-\dfrac{dE}{dx}\bigg\vert^{\text{soft}}=\dfrac{\pi e^2\nu^{-2\eps}}{v^2}F_{3+2\eps} \int_0^{\infty}  dq q^{3+2\eps} \int_{-v}^{v} dx x^2 (1-x^2/v^2)^{\eps}(1-x^2)^{\eps}
	\\
	\times \Bigg\lbrace \dfrac{  (1+\eps) m_{D,3+2\eps}^2   }{2\left(  q^2+m_D^2Q_1(x)\right)  ^2+\pi^2 x^2m_D^4/2 }
	+\dfrac{v^2-x^2}{1-x^2} \dfrac{ \left(  m_{D,3+2\eps}^2 -(e^2m^2/\pi^2)/(1-x^2)\right) }{\left(   2q^2+m_D^2Q_2(x)\right)  ^2+\pi^2 x^2   m_D^4 /4 }+\mathcal{O}(\eps)\Bigg\rbrace \ .
\end{gathered}
\ee
The theta function $\Theta(1-x^2)$ in the spectral functions does not affect the limits of integration for the energy transfer, since $v<1$. Furthermore, the momentum transfer integral contains an UV divergence for $q \gg eT$, which we regularize using DR. All remaining integrals are finite and can be computed analytically, they are given in Appendix \ref{B}. Hence, the leading order term of the soft contribution is
\be\label{soft_contribution_m=0_final}
\begin{gathered}
	-\dfrac{dE}{dx}\bigg\vert_{m=0}^{\text{soft}}= \dfrac{e^2m_{D,3+2\eps}^2}{16\pi}
	\bigg\lbrace  \left( -\dfrac{1}{\eps}+\ln \dfrac{\overline{\nu}^2}{m_D^2}\right) \left( \dfrac{1}{v}-\dfrac{1-v^2}{2v^2}\ln\dfrac{1+v}{1-v}\right)
	-\dfrac{2v}{3}-\dfrac{1}{v^2}\left( P(v)+A_{\text{soft}}(v)\right)\bigg\rbrace \ .
\end{gathered}
\ee
The extra finite pieces coming from the Debye mass in $d=3+2\eps$ dimensions, the function $P(v)$ given in Eq.\eqref{P(v)} and the term $-2v/3$ cancel exactly with those computed in the hard contribution. The function $A_{\text{soft}}(v)$ reads
\be
\begin{gathered}
	A_{\text{soft}}(v)=\int_{-v}^v dx \, x^2\bigg\lbrace\dfrac{1}{2}\ln \left( Q_1(x)^2+\dfrac{\pi^2x^2}{4}\right) 	+\dfrac{1}{4}\dfrac{v^2-x^2}{1-x^2}\ln \left( \dfrac{Q_2(x)}{4}^2+\dfrac{\pi^2x^2}{16}\right)
	\\
	+\dfrac{2Q_1(x)}{\pi x}\arccos\left( \dfrac{Q_1(x)}{\sqrt{Q_1(x)^2+\pi^2x^2/4}}\right)+\dfrac{v^2-x^2}{1-x^2}\dfrac{Q_2(x)}{\pi x}\arccos\left( \dfrac{Q_2(x)}{\sqrt{Q_2(x)^2+\pi^2x^2/4}}\right) \bigg\rbrace \ .
\end{gathered}
\ee
Focusing now on the mass dependent part of Eq.\eqref{Soft_contribution_intermediate_step} and computing the relevant integral in energy transfer  (see Eq.\eqref{Angular_integral_parte_hard_m^2}) we arrive to the final result for the mass correction to the soft contribution of the energy loss
\be\label{soft_contribution_m ^2_final}
\begin{gathered}
	-\dfrac{dE}{dx}\bigg\vert_{m^2}^{\text{soft}} \approx  \dfrac{e^2}{16\pi}\dfrac{e^2m^2}{2\pi^2}
	\left( -\dfrac{1}{\eps}+\ln \dfrac{\overline{\nu}^2}{m_D^2}\right) \left( \dfrac{3}{v}-\dfrac{v^2-3}{2v^2}\ln\dfrac{1+v}{1-v}\right)
	\ .
\end{gathered}
\ee
\section{Alternative computation of the soft contribution} 
\label{alternative_soft}	
In this section we will provide an alternative way of computing the soft contribution to the energy loss.  We will perform the computation starting from the general expression of the energy loss, given in Eq.\eqref{Energy_loss}, but using the amplitude of diagram Fig.\hyperref[diagram]{1.(b)}. The amplitude can be constructed following QED Feynman rules but replacing bare propagators by HTL resummed propagators in Coulomb gauge. Then, squaring the amplitude and performing the sum over spins of the involved particles in the process we find
\be\label{Amplitue_squared_dynamic_variables}
\begin{gathered}
	\dfrac{1}{2}\sum_{\text{spins}}\big\vert \mathcal{M}\big \vert ^2 =8e^4\nu^{6-2d} E^2\bigg\lbrace  
	\big\vert \Delta_L^d(Q)\big \vert ^2(E'/E+\bv'\cdot \bv+M^2/E^2)(E_k' E_{k}+\bk'\cdot\bk+m^2)
	\\
	+
	2\text{Re}[\Delta_L^d(Q)\Delta_T^d(Q)^*]
	\bigg[ (\bv'\cdot\bk_{\perp,\hat{q}}')E_k+ (\bv'\cdot\bk_{\perp,\hat{q}})E_{k}'+(\bv\cdot\bk_{\perp,\hat{q}}')E_k+(\bv\cdot\bk_{\perp,\hat{q}})E_{k}'\bigg] 
	\\
	+2\big\vert \Delta_T^d(Q)\big \vert ^2
	\bigg[ (\bv'\cdot\bk_{\perp,\hat{q}}')(\bv\cdot\bk_{\perp,\hat{q}})
	+(\bv'\cdot\bk_{\perp,\hat{q}})(\bv\cdot\bk_{\perp,\hat{q}}')+(\bv'_{\perp,\hat{q}}\cdot\bv_{\perp,\hat{q}})(E_{k}'E_k-\bk'\cdot\bk-m^2 )
	\\-\dfrac{d-1}{2}(E'/E-\bv'\cdot \bv-M^2/E^2)(E_{k}'E_k-\bk'\cdot\bk-m^2 )+(\bk_{\perp,\hat{q}}'\cdot \bk_{\perp,\hat{q}})(E'/E-\bv'\cdot \bv-M^2/E^2)\bigg]\bigg\rbrace \ .
\end{gathered}
\ee
Where $\bv=\p/E$ is the velocity of the heavy fermion and we also  defined $\bk_{\perp,\hat{q}}=-P_{\perp,\hat{q}}^{ij}k^i$ and $\bv_{\perp,\hat{q}}=-P_{\perp,\hat{q}}^{ij}v^i$, being $P_{\perp,\hat{q}}^{ij}=-(\delta^{ij}-\hat{q}^i\hat{q}^j)$ minus the transverse projector to $\hat{\q}=\q/q$.  We used primed variables for the outgoing particles. In the above expression, $\Delta^d_{L/T}(Q)$ denote the longitudinal and transverse components of HTL resumed propagators in $d$ dimensions respectively, assuming that the fermions in the plasma have mass $m$.
In order to simplify the expression for the amplitude we make use of the  exact kinematic relation 
\be \label{kinematic_soft}
\dfrac{M^2}{E^2}=\dfrac{E'}{E}-\bv'\cdot \bv+\dfrac{m^2}{E^2}-\dfrac{E_k'E_k-\bk'\cdot \bk}{E^2} \ .
\ee
Since the amplitude squared in Eq.\eqref{Amplitue_squared_dynamic_variables} is invariant under the interchange of $\bk \leftrightarrow \bk'$, we can antisymmetrize the thermal distribution functions \cite{Braaten:1989mz} i.e  replacing $n_F(E_k)[1-n_F(E_k')]$ by $ [n_F(E_k)-n_F(E_k')]/2$. Then we move to the transfer momentum variables, as we did in Sec.\ref{Hard} for the hard contribution. Eventually, we reach  the following expression for the soft contribution to the energy loss      
\be
\label{soft_contribution_transfer_variables}
\begin{gathered}
	-\dfrac{dE}{dx}\bigg\vert^{\text{soft}}= \dfrac{4\pi e^4\nu^{6-2d}}{v} \int \dfrac{d^dk}{(2\pi)^d } \int \dfrac{d^dq}{(2\pi)^d  } \int d\omega \omega 
	\dfrac{n_F(E_k)-n_F(E_k+\omega)}{E_k}
	\\
	\times \delta(t+2\omega E_k-2\bk\cdot\q)\delta\left( \omega-{\bv}\cdot{\q}-t/(2E)\right)
	\\
	\times  \bigg\lbrace  
	\big\vert \Delta_L^d(Q)\big \vert ^2\left( 1-\dfrac{\omega}{E}-\dfrac{\omega E_k-\bk\cdot \q}{2E^2}\right)(2E_k ^2+\omega E_k+\bk\cdot\q)
	+
	2\text{Re}[\Delta_L^d(Q)\Delta_T^{d}(Q)^*]
	(\bv\cdot\bk_{\perp,\hat{q}})\left(  2E_k+ \omega \right) 
	\\
	+\big\vert \Delta_T^d(Q)\big \vert ^2
	\bigg[ 2(\bv\cdot\bk_{\perp,\hat{q}})^2
	+\bv_{\perp,\hat{q}}^2(\omega E_k-\bk\cdot\q)
	+\left( \dfrac{\omega E_k-\bk\cdot \q}{E^2}\right) \left(  \bk_{\perp,\hat{q}}^2-\dfrac{d-1}{2}(\omega E_k-\bk\cdot \q ) \right)\bigg]\bigg\rbrace \ .
\end{gathered}
\ee
Now we introduce the average over velocities $\bv$ in $d$ dimensions using Eqs.\eqref{average_driections_v_a}-\eqref{average_driections_v_c}. We note that the piece that mixes the longitudinal and transverse component of the propagator vanishes in this process because it is transverse to $\q$. In addition, we eliminate all suppressed pieces in $E$ since we will work in the regime $T\ll E$. The integral over the angle $\theta\equiv \theta_{\bk,\q}$ is the same as that of the hard contribution, given in Eq.\eqref{angular_integration}. Then, moving to the dimensionless variable $x=\omega/q$ we can cast the soft contribution as
\be
\label{soft_contribution_after_average_velocities}
\begin{gathered}
	-\dfrac{dE}{dx}\bigg\vert^{\text{soft}}= \dfrac{2e^4\nu^{6-2d}}{v^2} C_d\int_0^\infty dkk^{d-2} \int_0^\infty dqq^{d-1} \int dx x \dfrac{n_F(E_k)-n_F(E_k+qx)}{E_k}
	\\
	\times \Theta\left( 1-\left[ \dfrac{q(x^2-1)}{2k}+\dfrac{x E_k}{k}\right] ^2\right) 
	\Theta(v^2-x^2)
	\bigg\lbrace  
	\big\vert \Delta_L^d(qx,q)\big \vert ^2\left( E_k ^2+qx E_k+\dfrac{q^2}{4}(x^2-1)\right) 
	\\
	+\big\vert \Delta_T^d(qx,q)\big \vert ^2
	\dfrac{1}{d-1} \left( v^2-x^2\right) \left(k^2-x^2E_k(qx+E_k)+qx E_k+\dfrac{q^2}{4}(x^2-1)\left( 3-x^2\right) \right) 
	\bigg\rbrace \times \mathcal{K}_d(x,q,k) \ . 
\end{gathered}
\ee
In the above expression, the momentum transfer is soft $q\sim eT$ and the momentum of the plasma constituents is hard $k\sim T$, which allows us to perform several approximations. The thermal distribution function can be expanded for $q\ll E_k$
\be\label{thermal_dist_expanded_low_q}
n_F(E_k+qx)=n_F(E_k)+qx \dfrac{dn_F(E_k)}{dE_k}+\mathcal{O}(q^2) \ .
\ee
In addition, the theta function in the second line can be simplified
\be
\Theta\left( 1-\left[ \dfrac{q(x^2-1)}{2k}+\dfrac{x E_k}{k}\right] ^2\right) \approx  \Theta\left( 1-\dfrac{x^2 E_k^2}{k^2}\right) \ .
\ee
Moreover, the function $\mathcal{K}_d(x,q,k)$ must be expanded for $q\ll E_k$ in order to be consistent with the HTL approximation
\be\label{F_d_aproximated}
\begin{gathered}
	\mathcal{K}_{d}(x,q,k) 
	\overset{}{=}(1-x^2/v^2)^{(d-3)/2}\left( 1-{x^2E_k^2}/{k^2}\right) ^{(d-3)/2}
	\left\lbrace 1+\dfrac{d-3}{2}\dfrac{xE_k}{k^2}\dfrac{1-x^2}{1-x^2E_k^2/k^2} q + \mathcal{O}(q^2) \right\rbrace \ .
\end{gathered}
\ee
Finally, we note that the terms $\sim q^n\big\vert \Delta_{L/T}^d(qx,q)\big \vert ^2$ for $n>3$ can be ignored, as they would contribute at higher order in the coupling constant.
Thus, the leading order pieces of the soft contribution to the energy loss are
\be
\label{soft_contribution_LO_pieces}
\begin{gathered}
	-\dfrac{dE}{dx}\bigg\vert^{\text{soft}}= \dfrac{2e^4\nu^{6-2d}}{v^2} C_d\int_0^\infty dkk^{d-2} \left( -\dfrac{1}{E_k}\dfrac{dn_F(E_k)}{dE_k}\right)  
    \\
    \times \int_0^\infty dqq^{d} \int dx x^2  (1-x^2/v^2)^{(d-3)/2}\left( 1-{x^2E_k^2}/{k^2}\right) ^{(d-3)/2}
	\\
	\times \Theta\left( 1-{x^2 E_k^2}/{k^2}\right) \Theta\left( v^2-x^2\right)
	\bigg\lbrace  
	\big\vert \Delta_L^{d}(qx,q)\big \vert ^2E_k ^2 
	+\big\vert \Delta_T^{d}(qx,q)\big \vert ^2
	\dfrac{v^2-x^2}{d-1} \left(k^2-x^2E_k^2 \right) 
	\bigg\rbrace 
	\ .
\end{gathered}
\ee
Explicit expressions for $\Delta_{L/T}^{d}$ are known for $m=0$. In Appendix \ref{C} we comment how the polarization tensors needed to build these propagators should be  computed for a generic value of the fermion mass. However, for a small fermion mass, and  at the order of accuracy we are going to compute, we will only need the resummed propagators at $m=0$.\\
Let us remark that the above expression may be valid for a generic mass $m$, as long as it does not surpass the plasma temperature. From  Eq.\eqref{soft_contribution_LO_pieces} we can reconstruct exactly Eq.\eqref{Soft_contribution_intermediate_step} in a very few steps. Setting $m=0$ and integrating by parts the distribution function, we easily arrive to
\be\label{soft_contribution_m=0}
\begin{gathered}
	-\dfrac{dE}{dx}\bigg\vert_{m=0}^{\text{soft}}= \dfrac{\pi e^2\nu^{3-d}}{v^2} m_{D,d}^2 F_d \int_0^{\infty} dqq^{d} \int_{-v}^{v} dx x^2 \left( 1-x^2/v^2\right)^{(d-3)/2} (1-x^2)^{(d-3)/2} \\	
    \times
	\bigg\lbrace 
	\dfrac{1}{2}\big\vert \Delta_L^d(q x,q)\big \vert ^2\dfrac{d-1}{2}+ \dfrac{1}{4}
	\big\vert \Delta_T^d(q x,q)\big \vert ^2 (v^2-x^2)(1-x^2)
	\bigg\rbrace \ .
\end{gathered}
\ee
Where we used the relation between the numerical constants defined through this manuscript $C_d=2\pi F_d^2$ and $m_{D,d}^2$ denotes the Debye mass squared in $d$ dimensions defined in Eq.\eqref{Debye_mass_in_d}. What remaings to be done is to insert the definition of the thermal HTL resummed propagators in $d=3+2\eps$ dimensions. They may be written as
\begin{subequations}
\begin{gather}\label{prop_T/L_m=0}
	\big\vert \Delta_{L}^{3+2\eps}(qx,q)\big \vert ^2=\dfrac{ 1}{\left( q^2+m_D^2Q_1(x)\right) ^2+\pi^2 m_D^4x^2/2 } +\mathcal{O}(\eps) \ ,
	\\\label{prop_T/L_m=0_2}
	\big\vert \Delta_{T}^{3+2\eps}(qx,q)\big \vert ^2=\dfrac{1}{(x^2-1)^2}\dfrac{ 1}{\left(  q^2+m_D^2Q_2(x)/2 \right) ^2+\pi^2 m_D^4x^2 /16} +\mathcal{O}(\eps) \ .
\end{gather}
\end{subequations}
When writing the longitudinal and transverse components of the propagators above, we did not include $\mathcal{O}(\eps)$ pieces, since they do not give rise to new UV divergences, and thus vanish in the limit $\eps\rightarrow 0$. We also did not include small mass corrections to the propagators, for the same reasons that we discarded the mass corrections in the denominators of the spectral functions (Eqs.\eqref{spectral_L}-\eqref{spectral_T}). Setting $d=3+2\eps$ everywhere and inserting the expression for the propagators of Eqs.\eqref{prop_T/L_m=0}-\eqref{prop_T/L_m=0_2}  in Eq.\eqref{soft_contribution_m=0} we get exactly Eq.\eqref{Soft_contribution_intermediate_step} for $m=0$.
\\
Now we show how to reproduce the remaining $\sim m^2$ term in Eq.\eqref{Soft_contribution_intermediate_step} from the more general expression Eq.\eqref{soft_contribution_LO_pieces}. 
In the regime $m\ll T\ll M$, the energy transferred to the plasma constituents is determined by $\Theta(x^2-v^2)$ rather than $\Theta(1-x^2k^2/E_k^2)$, as in this scenario the velocity of the heavy fermion is always smaller compared to the velocity of the fermionic particles in the plasma. Then, expanding Eq.\eqref{soft_contribution_LO_pieces} for small $m$ and keeping only pieces of $\mathcal{O}(m^2)$, we note that many pieces vanish after integrating by parts the thermal distribution functions, and the only non vanishing piece is 
\be\label{soft_contribution_m^2}
\begin{gathered}
	-\dfrac{dE}{dx}\bigg\vert_{m^2}^{\text{soft}}= -\dfrac{4\pi e^4m^2\nu^{6-2d}}{v^2} F_d^2\int_0^\infty dkk^{d-3}  \left(-\dfrac{dn_F}{dk}\right)
    \\
    \times
    \int_0^{\infty}dqq^{d} 
    \int_{-v}^{v} dx x^2 \left( 1-x^2/v^2\right)^{(d-3)/2} (1-x^2)^{(d-3)/2} 
    \dfrac{1}{d-1}
	\big\vert \Delta_T^d(x,q)\big \vert ^2 (v^2-x^2)
	\ .
\end{gathered}
\ee
Now the $k$ integral can be evaluated, yielding to
\be \label{k_integral_soft_alternative}
\begin{gathered}
	16 e^2m^2\nu^{3-d} \dfrac{F_d}{d-1} \int_{0}^{\infty} dk k^{d-3}\left(-\dfrac{dn_F}{dk} \right) =  
	\\
	=\dfrac{e^2m^2}{\pi^2}\left( 1+\left( -1-2\gamma_E-2\ln 2+2\ln\pi+\ln\dfrac{T^2}{\overline{\nu}^2}\right)\eps+\mathcal{O}(\eps^2) \right) 
\end{gathered}
\ee
The finite pieces $\sim\eps$ in the second line can be ignored at the order we are working and, as we pointed below Eq.\eqref{Integral_k_hard_m^2_II}, the logarithms $\ln(T^2/\overline{\nu}^2)$ cancel when the hard and soft contribution are added. Thus, we may write Eq.\eqref{soft_contribution_m^2} as
\be\label{soft_contribution_m^2_int_by_parts}
\begin{gathered}
	-\dfrac{dE}{dx}\bigg\vert_{m^2}^{\text{soft}}= \dfrac{\pi e^2\nu^{-2\eps}}{v^2} F_{3+2\eps} \int_0^{\infty}dqq^{3+2\eps} \int_{-v}^{v} dx x^2 \left( 1-x^2/v^2\right)^{\eps} (1-x^2)^{\eps} \Theta\left( 1-x^2\right)
	\\	
	\times \dfrac{1}{4} \dfrac{v^2-x^2}{(1-x^2)^2}  \dfrac{ (-e^2m^2/\pi^2)}{\left (  q^2+m_D^2Q_2(x)/2 \right)^2+\pi^2 m_D^4x^2 /16}  
	\ .
\end{gathered}
\ee
This is the same expression we found for the mass  correction of the soft sector in
Eq.\eqref{Soft_contribution_intermediate_step}, as expected.

\section{  Results and discussion}
\label{Final}

\begin{figure}
	\centering
	\DeclareGraphicsExtensions{.pdf}
	\includegraphics[scale = 1 ]{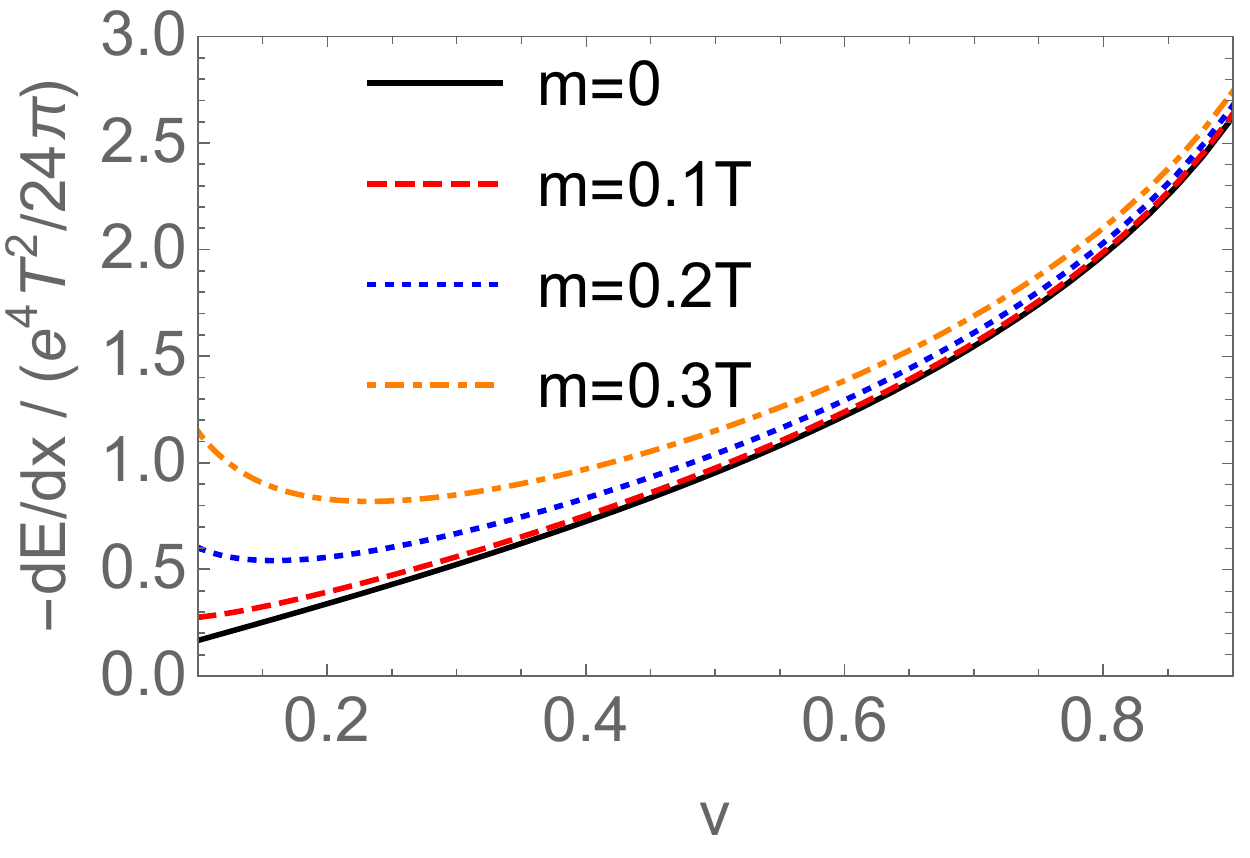}
	\caption{Values of the collisional energy loss in QED for the different values of the mass of the fermion constituents of the plasma. The black line corresponds to the massless case, the red dashed line for $m=0.1 \,T$, the blue dashed line for $m=0.2 \, T$, and the orange dot-dashed line $m=0.3 \, T$.}
	\label{plotresultados}
\end{figure}

The final expression of the collisional energy loss
is obtained after adding the hard and soft contributions. Then, the poles $1/\eps$ and 
the dependence on the renormalization scale $\nu$ cancel out. For $m=0$
we reach to the same expression first found by Braaten and Thoma \cite{Braaten:1991jj} in the regime $E \ll M^2/T$
\begin{equation} \label{complete_result_m=0}
		-\dfrac{dE}{dx}\bigg\vert_{\text{BT}} = 
		\dfrac{e^2 m^2_{D}}{8\pi}   \left( \dfrac{1}{v}-\dfrac{1-v^2}{2v^2}\ln\dfrac{1+v}{1-v}\right) 
        \left\lbrace
        \ln  \dfrac{E}{M}  +\ln \dfrac{1}{e}+ A(v) \right\rbrace \ .
\end{equation}
The extra finite pieces generated in the computation of the hard and
soft contributions using DR  also cancel out, so that the function $A(v)$ is the same as that using a cutoff for the computation.
 Note that Eq.\eqref{complete_result_m=0} is
not valid for neither the $v\rightarrow 0$ or $v \rightarrow 1$ limits, because of the kinematical constraints  used in the evaluation of the
momentum intregrals. In particular,  we assumed that the velocity of the heavy fermion is always smaller than that of the plasma constituents. Taking into account those limits can
be  done after a proper modification of the kinematical constraints,  see Refs.\cite{Braaten:1991jj,Peigne:2007sd}. Further,  in the case $v \rightarrow 1$  Compton scattering also contributes at the same order as the one here computed \cite{Peigne:2008wu}.
\\
The fermion mass correction to the above result is also obtained after adding the corresponding hard and soft contributions we computed. Then,  the pole and  the dependence on the renormalization scale also cancel out, and to leading logarithmic accuracy we obtain 
for $m \ll T \ll M$  
%
%
%
\begin{equation}
		-\dfrac{dE}{dx}\bigg\vert = 		-\dfrac{dE}{dx}\bigg\vert_{\text{BT}} + \dfrac{e^4m^2}{16\pi^3}\left( \dfrac{3}{v}-\dfrac{v^2-3}{2v^2}\ln\dfrac{1+v}{1-v}\right)\ln\left(\frac{1}{e}\right)  \ .
 \end{equation}
 We have not computed mass corrections beyond logarithmic accuracy, as 
most likely the genuine perturbative corrections to the energy loss are more relevant  in the regime where our assumptions are valid. \\
We have represented in Fig.~\ref{plotresultados} the value of the collisional energy loss for different values of the fermion mass, so as to estimate how relevant its effect could be. We  note that 
the effects of a fermion mass seem to be quite relevant already for values of $m = 0.3 \,T$. It might thus seem interesting to evaluate also the energy loss for
values of the fermion mass close to $T$, where our approximations are not valid.
Note that  the assumption $m \ll T$ allows us to compute the collisional energy loss analytically, which is always a good initial step to assess the effect we are considering. We have provided all the ingredients to carry out the computation for values of $m$ getting close to $T$, but we defer the careful study of that case for future projects, as it requires a much more detailed analysis. The main difficulty for such a computation 
is the evaluation of the soft sector,  as one should generalize the Braaten-Pisarski resummation program in the presence of massive fermions. The explicit form of the photon polarization tensor needed in that case to construct the resummed propagators is given in Appendix
\ref{C}.

Although our computation was initially performed for a QED plasma, it can be  extended to QCD. In fact, retaining mass fermion corrections in QCD can be fully justified, and it was our ultimate motivation. For instance, in the context of heavy ion collisions, it may be reasonable to disregard the masses of up and down quarks, but neglecting the mass of the strange quark may not be such a good approximation.
The contribution of the mass effects  in QCD can be obtained from the corresponding QED calculation
at logarithmic accuracy by simple substitution, while further corrections would be needed beyond this accuracy. Note that the mass corrections
to the HTL gluon polarization tensor are the same as those of the photon polarization in 
QED, only  some color
and flavor factors have to be taken into account. Similarly, in the evaluation of the scattering matrix element of a heavy quark with a light quark in the $t$-channel, the one that is IR sensitive, only some color and flavor factors are to be considered.  At leading logarithmic accuracy the  corrections associated to a massive quark  to the QCD collisional energy loss are also given by the QED result,  replacing $e^2$ by 
$g^2$, the strong coupling constant, and taking into account the  factor $2/3$. More explicitly
\begin{equation}
		\dfrac{g^4m^2}{24 \pi^3}\left( \dfrac{3}{v}-\dfrac{v^2-3}{2v^2}\ln\dfrac{1+v}{1-v}\right)\ln\left(\frac{1}{g}\right)  \ .
 \end{equation}
We have evaluated the impact of including the strange quark mass
corrections to the collisional energy loss of a charm and bottom quark (see Figs.2 and 3 of Ref.~\cite{Braaten:1991we}), when $T=250$ MeV, and assuming that the strong fine structure constant is 
$\alpha_s= 0.2$. Taking the strange quark mass as $m = 100$ MeV, so $m = 0.4 \,T$, we note that in this case the mass corrections are in the 1 to 2 percent level. The effect is
certainly not as large as in QED because of two reasons. First, the gauge coupling constant is larger in QCD, and second, the contribution of one parton, no matter whether it is massless or massive, can never be too big as compared to the contribution associated to all partons.
In any case, the mass effects here discussed are there and we have provided the tools to assess them.

\section*{Acknowledgements} 
\label{ack}		
We thank Joan Soto for useful discussions.
This work was supported by Ministerio de Ciencia, Investigaci\'on y Universidades (Spain) under the project PID2019-110165GB-I00 (MCI/AEI/FEDER, UE), 
by Generalitat de Catalunya by the project 2021-SGR-171 (Catalonia). This work was also partly supported by the Spanish program Unidad de Excelencia
 Maria de Maeztu CEX2020-001058-M, financed by MCIN/AEI/10.13039/501100011033.

\appendix

\section{Integrals of the hard contribution}\label{A}	
\renewcommand{\theequation}{A.\arabic{equation}}
\setcounter{equation}{0}

When computing the hard contribution to the energy loss, we encounter three types of momentum transfer integrals in the region $0<q<q_{\text{in}}$, see Eq.\eqref{Hard_contribution_in_d_leading_order}. When computed in $d=3+2\eps$, all of them give rise to Appell $F_1$ hypergeometric functions, due to the particular dependence on momentum transfer $q$ of the function $\mathcal{K}_d(x,q,k)$ defined in Eq.\eqref{function}. Explicitly,  we compute the general formula
\be \label{momentum_transfer_integral_hard_master_formula}
\begin{gathered}
	\nu^{3-d}\int_0^{L}dq q^{d-n} \left( 1-\left[ \dfrac{q(x^2-1)}{2k}+\dfrac{x E_k}{k}\right]^2\right)^{\frac{(d-3)}{2}}  
	=\dfrac{L^{d-n+1}}{d-n+1} \left( 1-\dfrac{x^2E_k^2}{k^2}\right)^{\frac{(d-3)}{2}} 
	\\
	\times
	F_1\left( 1+d-n,-\dfrac{d-3}{2},-\dfrac{d-3}{2},d-n+2,\dfrac{1-x^2}{k+xE_k}L,-\dfrac{1-x^2}{k-xE_k}L\right)  \ ,
\end{gathered}  
\ee
where $L\equiv 2(k-vE_k )/(1-v^2)$ and $F_1(a;b_1,b_2;c;z_1,z_2)$ is the so called Appell hypergeometric function of the first kind. For $n=5, 4, 3$ we obtain the required integrals for the evaluation of the hard contribution. The above results can be simplified expressing $F_1$ by its infinite series representation 
\be \label{Apell_series}
F_1(a;b_1,b_2;c;z_1,z_2)=\sum_{m=0}^{\infty} \sum_{n=0}^{\infty} \dfrac{(a)_{m+n} (b_1)_{m}(b_2)_{n}}{m! \ n! \ (c)_{m+n}} z_1^m z_2^n \ ,
\ee
where $(\alpha)_n= \Gamma(n+\alpha)/\Gamma(\alpha)$ is a Pochhammer symbol, and realizing that in $d=3+2\eps$ only the first terms of the series are non-vanishing. Following this procedure, we obtain Eq.\eqref{q_integral_hard_finite} and Eq.\eqref{q_integral_hard_IR} setting $n=5$ and $n=4$ in Eq.\eqref{momentum_transfer_integral_hard_master_formula} respectively. As stated in Sec.\ref{Hard} the last case $n=3$ is not necessary for the evaluation of the hard contribution. In order to see it, we use Eq.\eqref{Apell_series} to write the Appell function as
\be
F_1\left( d -2,-\dfrac{d-3}{2},-\dfrac{d-3}{2},d-1,\dfrac{k-vE_k}{k+xE_k}\dfrac{1-x^2}{1-v^2},\dfrac{k-vE_k}{k-xE_k}\dfrac{1-x^2}{1-v^2} \right)= 1+\mathcal{O}\left( d-3\right) \ .
\ee
Note that Eq.\eqref{momentum_transfer_integral_hard_master_formula} is finite in three spatial dimensions when $n=3$. Then, since in $d=3$ the measure of Eq.\eqref{Hard_contribution_in_d_leading_order} is even in $x$, the corresponding terms vanish due to antisymmetry in $x$. Let us now write the results needed for the evaluation of the hard contribution at $m=0$ i.e Eq.\eqref{Hard_leading_order_m=0}. The integral in energy transfer gives
\be \label{energy_transfer_integral_m=0}
\begin{gathered}
	\int_{-v}^{v} dx \left( 1-{x^2}/{v^2}\right)^{(d-3)/2}\left( 1-x^2\right)^{(d-3)/2}\bigg\lbrace
	3x^2+\dfrac{x^2(1-v^2)}{x^2-1}+ (d-3) x^2
	\bigg\rbrace =
	\\
	=2v^2\left\lbrace \left( \dfrac{1}{v}+\dfrac{v^2-1}{2v^2}\ln\dfrac{1+v}{1-v}\right)+\eps\dfrac{2v}{3} +\eps \dfrac{P(v)}{v^2}+\mathcal{O}(\eps^2)\right\rbrace  \ .
\end{gathered}
\ee
Where $P(v)$ was previously defined in Eq.\eqref{P(v)}. The result for the $k$ integral may be written as
\be \label{k_integral_m=0}
\begin{gathered}
	2 e^4\nu^{3-d}  C_{d}  \int_0^\infty dk \ k^{d-2} \left(\frac{2k}{(1+v)\nu}\right)^{d-3} n_F(k)= 
	\\
	= \dfrac{e^2m_{D,3+2\eps}^2}{16\pi }\left\lbrace 1+\left[ 1+\gamma_E+\dfrac{\zeta'(2)}{\zeta(2)}+\ln 2+\dfrac{1}{2}\ln \dfrac{4T^2}{(1+v)^2\overline{\nu}^2} \right]2\eps +\mathcal{O}(\eps^2) \right\rbrace \ .
\end{gathered}
\ee
When writing the above result, we used the Debye mass in $d=3+2\eps$ dimensions defined in Eq.\eqref{Debye_mass_in_d}. Multiplying the series Eq.\eqref{energy_transfer_integral_m=0} by Eq.\eqref{k_integral_m=0} we eventually reach the result for the hard contribution to the energy loss at $m=0$ i.e Eq.\eqref{hard_final_m=0}.

Let us now discuss how to properly regularize the momentum integrals for the mass corrections of the hard contribution, i.e the $k$ integrals of Eq.\eqref{Hard_contribution_m^2_after_q_integrals}. Explicitly, moving to the dimensionless variable $y=k/T$ and setting $d=3+2\eps$, the IR integral to evaluate is
\be\label{k_integral_hard_appendix}
\begin{gathered}
	e^4 m^2 C_{3+2\eps}\left( \dfrac{T}{\nu}\right)^{2\eps} \int_0^{\infty} dy \ y^{-1+2\eps} \left(\frac{2T}{(1+v)\nu}y\right)^{2\eps} n_F(y)
	\ .
\end{gathered}
\ee
In DR, the small parameter $\eps$ acts as a regulator of the divergence, playing a similar role as a cut-off would do. In order to properly regularize Eq.\eqref{k_integral_hard_appendix} we must distinguish the regulator of the IR divergence in momentum transfer from the regulator of the IR divergence in $k$, that we denote as $\eps_k$. So, instead of Eq.\eqref{k_integral_hard_appendix} we need to evaluate
\be\label{Integral_k_hard_m^2_step0}
\begin{gathered}
	e^4 m^2 C_{3+2\eps}\left( \dfrac{T}{\nu}\right)^{2\eps} \int_0^{\infty} dy \ y^{-1+2\eps_k} \left(\frac{2T}{(1+v)\nu}y\right)^{2\eps} n_F(y)
	=
	\\
	=e^4 m^2 C_{3+2\eps}\left( \dfrac{T}{\nu}\right)^{2\eps}\left( \dfrac{2T}{(1+v)\nu}\right)^{2\eps} (1-2^{1-2\eps-2\eps_k})\Gamma(2\eps+2\eps_k)\zeta(2\eps+2\eps_k) \ .
\end{gathered}
\ee
Taking the limit $\eps_k\rightarrow0$ above and then expanding for $\eps\rightarrow 0$ we get Eq.\eqref{Integral_k_hard_m^2_II}. In this way, the piece that was initially IR divergent in $k$  also has a contribution to the pole in $\eps$ i.e the regulator of the IR divergence in momentum transfer. Applying the same procedure we can compute the other necessary integral in momentum of Eq.\eqref{Hard_contribution_m^2_after_q_integrals}, which gives
\be\label{Integral_k_hard_m^2_I}
\begin{gathered}
	e^4 m^2 \nu^{3-d}C_{d}\int_0^{\infty} dk \ k^{d-3} \left(\frac{2k}{(1+v)\nu}\right)^{d-3} \dfrac{1}{2}\dfrac{dn_F}{dk}=
	\\
	=\dfrac{e^2}{16\pi}\dfrac{e^2m^2}{2\pi^2}\left\lbrace -1-\left( -2\gamma_E+2\ln\pi +\ln \dfrac{T^2}{\overline{\nu}^2}+\ln \dfrac{4T^2}{(1+v)^2\overline{\nu}^2}\right)\eps+\mathcal{O}(\eps^2) \right\rbrace \ .
\end{gathered}
\ee
Although the result of the above integral is finite, intermediate steps in the computation require the evaluation of an IR divergent integral so that the same method for regularizing Eq.\eqref{Integral_k_hard_m^2_II} must be used. If we do not distinguish between the regulators of the different divergences when evaluating the momentum integrals of the mass corrections, we would be unable to cancel the IR divergence in momentum transfer of the hard contribution with the UV divergence in momentum transfer of the soft contribution. 

\section{Integrals of the soft contribution}\label{B}	
\renewcommand{\theequation}{B.\arabic{equation}}
\setcounter{equation}{0}

The logarithmic ultraviolet divergence encountered in the soft region of the computation 
can be computed using DR regularization. One only needs to evaluate in
$d=3 + 2 \eps$
\cite{Carignano:2021mrn}
\be\label{q_integral_soft_lon}
\begin{gathered}
	\nu^{3-d}\int_0^\infty dq \dfrac{ q^d}{ \left (  q^2+m_D^2a \right) ^2+m_D^4b^2 }=
 \\
	=-\dfrac{1}{2} \left\lbrace \dfrac{1}{\eps}-\ln\left( \dfrac{\nu^2}{m_D^2}\right) +\dfrac{1}{2}\ln \left(a^2+b^2\right) +\dfrac{a}{b}\arccos\left( \dfrac{a}{\sqrt{a^2+b^2}}\right) \right\rbrace +\mathcal{O}(\eps) \ .
\end{gathered}
\ee
\section{HTL Polarization tensor for generic fermion mass $m$}\label{C}	
\renewcommand{\theequation}{C.\arabic{equation}}
\setcounter{equation}{0}

We have previously mentioned that Eq.\eqref{Hard_contribution_after_q_integral} and Eq.\eqref{soft_contribution_LO_pieces} should be valid for the evaluation of the hard and soft contribution to the energy loss respectively, if the fermions in the plasma have  mass $m$, which is assumed to be at most of order $T$. 
However, the computation of the soft contribution with generic mass demands that the HTL resummation technique has to be done assuming massive fermions.
We give here the expressions of the HTL polarization tensors in such a case. These can be evaluated, for example, in the real time formalism of thermal field theory, following the same steps as in 
Ref.~\cite{Carignano:2017ovz} but with massive particles.  After carrying out the integral in frequency one then arrives to the expression 
\be
\begin{gathered}
\Pi^{\mu\nu}(L)= e^2\nu^{3-d}\int\dfrac{d^dq}{(2\pi)^d} \dfrac{1-2n_F(E_q)}{E_q}\bigg\lbrace\dfrac{2E_qv^\mu v^\nu-v^\mu L^\nu-v^\nu L^\mu+g^{\mu\nu}({v}\cdot{L})}{{v}\cdot{L}-L^2/2E_q+ i\text{sgn}(E_q-l_0) 0^+}
\\
-\dfrac{2E_q\tilde{v}^\mu \tilde{v}^\nu+\tilde{v}^\mu L^\nu+\tilde{v}^\nu L^\mu-g^{\mu\nu}({\tilde{v}}\cdot{L})}{{\tilde{v}}\cdot{L}+L^2/2E_q+ i\text{sgn}(E_q+l_0) 0^+}
\bigg\rbrace \ .
\end{gathered}
\ee
Here $L=(l_0,\l)$ is the  external photon momentum, $v^\mu=(1,\q/E_q)$ and 
$\tilde{v}^\mu =(1,-\q/E_q) $ are
the velocity of the massive fermions and antifermions, respectively, with $E_q=\sqrt{q^2+m^2}$ and $\text{sgn}(x)$ is the sign function.
Under the assumption that external momenta obeys $L \ll Q$, after a change of variables to express the antiparticle contribution as the particle one, it is possible to approximate the expression to find 
\be
\begin{gathered}
\Pi^{\mu\nu}(L) \approx 2 e^2\nu^{3-d}\int\dfrac{d^dq}{(2\pi)^d} \dfrac{1-2n_F(E_q)}{E_q} \left(g^{\mu \nu} - \frac{ v^\mu L^\nu+v^\nu L^\mu}{{v}\cdot{L}}
+\frac{ L^2 v^\mu v^\nu}{({v}\cdot{L})^2} \right)
\ .
\end{gathered}
\ee
Note that in the limit $m=0$, one reproduces the correct well-known limit for the HTL.
This final result is the one that one would obtain from classical transport theory. Unfortunately, there is not a simple analytical expression of the polarization tensor for a generic value $m$.
The generalization of the Braaten-Pisarski resummation program in this case would most likely to be implemented numerically.


\begin{thebibliography}{99}


\bibitem{dEnterria:2009xfs}
D.~d'Enterria,
Landolt-Bornstein \textbf{23}, 471 (2010)
doi:10.1007/978-3-642-01539-7\_16
[arXiv:0902.2011 [nucl-ex]].

\bibitem{Casalderrey-Solana:2007knd}
J.~Casalderrey-Solana and C.~A.~Salgado,
Acta Phys. Polon. B \textbf{38}, 3731-3794 (2007)
[arXiv:0712.3443 [hep-ph]].

\bibitem{Majumder:2010qh}
A.~Majumder and M.~Van Leeuwen,
Prog. Part. Nucl. Phys. \textbf{66}, 41-92 (2011)
doi:10.1016/j.ppnp.2010.09.001
[arXiv:1002.2206 [hep-ph]].

\bibitem{Qin:2015srf}
G.~Y.~Qin and X.~N.~Wang,
Int. J. Mod. Phys. E \textbf{24}, no.11, 1530014 (2015)
doi:10.1142/S0218301315300143
[arXiv:1511.00790 [hep-ph]].

\bibitem{Bjorken:1982tu}
J.~D.~Bjorken,
FERMILAB-PUB-82-059-THY.

\bibitem{Braaten:1991jj}
E.~Braaten and M.~H.~Thoma,
Phys. Rev. D \textbf{44}, 1298-1310 (1991)
doi:10.1103/PhysRevD.44.1298

\bibitem{Braaten:1991we}
E.~Braaten and M.~H.~Thoma,
Phys. Rev. D \textbf{44}, no.9, R2625 (1991)
doi:10.1103/PhysRevD.44.R2625
				
\bibitem{Pisarski:1988vd}
R.~D.~Pisarski,
Phys. Rev. Lett. \textbf{63}, 1129 (1989)
doi:10.1103/PhysRevLett.63.1129
				
\bibitem{Braaten:1989mz}
E.~Braaten and R.~D.~Pisarski,
Nucl. Phys. B \textbf{337}, 569-634 (1990)
doi:10.1016/0550-3213(90)90508-B
				
				
\bibitem{Peigne:2007sd}
S.~Peigne and A.~Peshier,
Phys. Rev. D \textbf{77} (2008), 014015
doi:10.1103/PhysRevD.77.014015
[arXiv:0710.1266 [hep-ph]].

\bibitem{Peigne:2008nd}
S.~Peigne and A.~Peshier,
Phys. Rev. D \textbf{77}, 114017 (2008)
doi:10.1103/PhysRevD.77.114017
[arXiv:0802.4364 [hep-ph]].

\bibitem{Ghiglieri:2020dpq}
J.~Ghiglieri, A.~Kurkela, M.~Strickland and A.~Vuorinen,
Phys. Rept. \textbf{880}, 1-73 (2020)
doi:10.1016/j.physrep.2020.07.004
[arXiv:2002.10188 [hep-ph]].

\bibitem{Manuel:2016wqs}
C.~Manuel, J.~Soto and S.~Stetina,
Phys. Rev. D \textbf{94}, no.2, 025017 (2016)
[erratum: Phys. Rev. D \textbf{96}, no.12, 129901 (2017)]
doi:10.1103/PhysRevD.94.025017
[arXiv:1603.05514 [hep-ph]].

\bibitem{Carignano:2017ovz}
S.~Carignano, C.~Manuel and J.~Soto,
Phys. Lett. B \textbf{780}, 308-312 (2018)
doi:10.1016/j.physletb.2018.03.012
[arXiv:1712.07949 [hep-ph]].

\bibitem{Carignano:2021zhu}
S.~Carignano and C.~Manuel,
Phys. Rev. D \textbf{104}, no.5, 056031 (2021)
doi:10.1103/PhysRevD.104.056031
[arXiv:2107.03655 [hep-ph]].

\bibitem{Carignano:2019ofj}
S.~Carignano, M.~E.~Carrington and J.~Soto,
Phys. Lett. B \textbf{801}, 135193 (2020)
doi:10.1016/j.physletb.2019.135193
[arXiv:1909.10545 [hep-ph]].

\bibitem{Gorda:2022fci}
T.~Gorda, A.~Kurkela, J.~\"Osterman, R.~Paatelainen, S.~S\"appi, P.~Schicho, K.~Sepp\"anen and A.~Vuorinen,
Phys. Rev. D \textbf{107}, no.3, 036012 (2023)
doi:10.1103/PhysRevD.107.036012
[arXiv:2204.11279 [hep-ph]].


\bibitem{Ekstedt:2023anj}
A.~Ekstedt,
[arXiv:2302.04894 [hep-ph]].


\bibitem{Comadran:2021pkv}
M.~Comadran and C.~Manuel,
Phys. Rev. D \textbf{104} (2021) no.7, 076006
doi:10.1103/PhysRevD.104.076006
[arXiv:2106.08904 [hep-ph]].
				
\bibitem{Carignano:2021mrn}
S.~Carignano and C.~Manuel,
Phys. Rev. D \textbf{103} (2021) no.11, 116002
doi:10.1103/PhysRevD.103.116002
[arXiv:2103.02491 [hep-ph]].
				



\bibitem{LeBellac}
M. Le Bellac,  {\it Thermal Field Theory}, Cambridge University Press, Cambridge 1996.

\bibitem{Peigne:2008wu}
S.~Peigne and A.~V.~Smilga,
Phys. Usp. \textbf{52}, 659-685 (2009)
doi:10.3367/UFNe.0179.200907a.0697
[arXiv:0810.5702 [hep-ph]].
     
\end{thebibliography}
\end{document}